\documentclass[conference]{IEEEtran}
\IEEEoverridecommandlockouts

\usepackage{cite}
\usepackage{amsmath,amssymb,amsfonts}
\usepackage{graphicx}
\usepackage{textcomp}
\usepackage{xcolor}

\usepackage{amssymb}
\usepackage{amsmath}
\usepackage{mathdots}
\usepackage{enumerate}
\usepackage{tikz}
\usepackage{mleftright}
\usepackage{setspace}
\usepackage{bm}

\usepackage{afterpage}
\usepackage{caption}
\usepackage{graphicx}
\usepackage{float}
\usepackage{subfigure}
\usepackage{booktabs}
\usepackage{multirow}
\usepackage{url}            
\usepackage[hidelinks]{hyperref}

\usepackage{algorithm}
\usepackage{algpseudocode}

\usetikzlibrary{decorations.fractals,spy,positioning, fit, calc}

\newcommand{\annotdim}[1]{{\footnotesize\color{gray} #1}}

\definecolor{gray}{rgb}{0.5, 0.5, 0.5}

\newcommand{\calN}{\mathcal{N}}
\newcommand{\calP}{\mathcal{P}}
\newcommand{\Fig}[1]{Fig.~\ref{#1}}

\newcommand{\R}{\mathbb{R}}

\DeclareMathOperator{\sgn}{sgn}

\makeatletter
\DeclareFontFamily{U}{MnSymbolA}{}
\DeclareSymbolFont{MnSyA}{U}{MnSymbolA}{m}{n}
\DeclareFontShape{U}{MnSymbolA}{m}{n}{
    <-6> MnSymbolA5
    <6-7> MnSymbolA6
    <7-8> MnSymbolA7
    <8-9> MnSymbolA8
    <9-10> MnSymbolA9
    <10-12> MnSymbolA10
    <12-> MnSymbolA12}{}
\DeclareMathSymbol{\smallrightarrow}{\mathrel}{MnSyA}{0}
\DeclareMathSymbol{\smallleftarrow}{\mathrel}{MnSyA}{2}
\DeclareMathSymbol{\smallleftrightarrow}{\mathrel}{MnSyA}{16}
\newcommand{\smallrightarrowfill@}{\arrowfill@\relbar\relbar\smallrightarrow}
\newcommand{\smallleftarrowfill@}{\arrowfill@\smallleftarrow\relbar\relbar}
\newcommand{\smallleftrightarrowfill@}
{\arrowfill@\smallleftarrow\relbar\smallrightarrow}
\renewcommand{\overrightarrow}{\mathpalette{\overarrow@\smallrightarrowfill@}}
\renewcommand{\overleftarrow}{\mathpalette{\overarrow@\smallleftarrowfill@}}
\renewcommand{\overleftrightarrow}
{\mathpalette{\overarrow@\smallleftrightarrowfill@}}

\makeatother

\providecommand{\msgf}[2]{\protect\overrightarrow{#1}_{\mspace{-3mu}#2}} 
\providecommand{\msgb}[2]{\protect\overleftarrow{#1}_{\mspace{-3mu}#2}} 
\def\BibTeX{{\rm B\kern-.05em{\sc i\kern-.025em b}\kern-.08em
    T\kern-.1667em\lower.7ex\hbox{E}\kern-.125emX}}
\begin{document}

\title{A Single-Parameter Factor-Graph Image Prior\\
}

\author{\IEEEauthorblockN{Tianyang Wang, Ender Konukoglu, Hans-Andrea Loeliger}
\IEEEauthorblockA{
Dept.\ of Information Technology and Electrical Engineering \\
ETH Zurich, Switzerland \\
tianwang@isi.ee.ethz.ch, ender.konukoglu@vision.ee.ethz.ch, loeliger@isi.ee.ethz.ch}
}

\maketitle

\begin{abstract}
We propose a novel piecewise smooth image model 
with piecewise constant local parameters that are automatically 
adapted to each image.
Technically, the model is formulated in terms
of factor graphs with NUP (normal with unknown parameters) priors, 
and the pertinent computations amount to iterations of conjugate-gradient steps 
and Gaussian message passing.
The proposed model and algorithms are demonstrated 
with applications to denoising and contrast enhancement.
\end{abstract}


\section{Introduction}

Modern methods for image processing are mostly based on deep neural networks.
However, deep neural networks need to be trained on large data bases 
and are prone to complex hallucinations,
which limits their attractivity in some applications. 
Moreover, adapting such methods to different distributions and imaging tasks 
usually requires the networks to be retrained.

By constrast, explicit prior models need no training, 
and their inductive bias 
is more transparent and justifiable.
It is therefore 
desirable to improve the performance and flexibility of such models
even the era of deep learning.

One class of such models (or regularizers) simply penalizes 
the difference between neighboring pixels 
with a suitable cost function.
The most popular such priors/regularizers 
are based on the L1 norm 
and favor piecewise constant images;
this includes 
total-variation (TV) regularization \cite{rudin1992nonlinear, chambolle2004algorithm}, 
total generalized variation (TGV) \cite{bredies2010total}, 
and structure tensor total variation (STV) \cite{lefkimmiatis2015structure}.

Another class of such models penalizes second-order differences 
\cite{yezzi1998modified,zhu2012image,zhong2020minimizing}.
However, the method of \cite{zhong2020minimizing} 
requires three global parameters to be tuned.
%

Better results have been reported for penalizing pixel differences 
with cost functions that are 
convex for small pixel differences and concave for large pixel differences \cite{ma2022smoothed};
the convex part prefers smoothly varying areas while the concave part allows sharp edges.
However, 
the prior proposed in \cite{ma2022smoothed} 
has two global parameters that need to be tuned 
and cannot easily be adapted to different parts of an image.

In this paper, we propose a novel image model 
that is more flexible, but also more complex, than such priors/regularizers 
in the prior literature.
Like the model of \cite{ma2022smoothed},
the proposed model favors piecewise smooth images,
but unlike \cite{ma2022smoothed}, 
the proposed model works with many local parameters.
Despite this additional model complexity,
these parameters are automatically adapted to each image.
There is 
only one global parameter, 
which is interpretable and determines the tradeoff
between matching the data and fitting the model;
it is the only parameter that must be set by the user.

Technically, the proposed model uses both NUP (normal with unknown parameters) priors 
\cite{loeliger2018factor}
as well as a method proposed in \cite{loeliger2023nup} 
to estimate piecewise constant scale factors.
With these ideas, the model fitting 
can be reduced to 
variations of iteratively reweighted least squares 
using both conjugate gradient steps and Gaussian message passing.
The use of all these techniques in concert 
may be of interest beyond the specific model of this paper.

The viability of the proposed model and algorithms are demonstrated 
with applications to denoising and contrast enhancement,
and with additional applications and examples in the appendix of the arXiv version (arXiv:2601.08749).

\section{Proposed Model and Estimation Algorithm}

We will work with images consisting of
pixels that are arranged in a rectangular grid.
For ease of notation, 
we use only single indices 
for pixel values,
e.g., $y_n\in\R$ for grayscale and $y_n\in\R^3$ for color images.

In this paper, we assume a very simple observation model 
where a latent pixel value $Y_n$
is related to an observed noisy pixel value $\breve Y_n = \breve y_n$ 
by 
\begin{equation} \label{eqn:ObsYn}
\breve Y_n = Y_n + Z_n,
\end{equation}
where $Z_n$ is white Gaussian noise with variance $\sigma_Z^2$ for gray scale 
and with covariance matrix $\sigma_Z^2 I\in \R^{3\times 3}$ for color images.
The assumed scalar noise variance $\sigma_Z^2>0$ 
is the mentioned global parameter 
that determines the tradeoff between 
fitting the data and matching the model.

Experiments with, and variations of, this setting
will be discussed in Section~\ref{sec:Experiments}.
Using the proposed model with nontrivial observations models 
is certainly possible, but will not be considered in this paper.

The basic idea of the proposed model 
is to encourage the latent pixel values $Y_n$ to be (approximately) piecewise smooth,
with 
unsupervised partitioning and 
locally adapted parameters,
as will be detailed in Sections \ref{sec:BasicModel} to~\ref{sec:BasicAlgorithm}.

An augmentation of the model
(described in Section~\ref{sec:Phase2})
aims at getting back some details (e.g., texture) into the smooth parts 
that might have been oversmoothed by the basic model.


\subsection{Basic Model}
\label{sec:BasicModel}

For every row and every column of the image, 
there is a state space model with latent state $X_n$ 
(comprising the pixel intensity and the slope)
evolving along the row or column, respectively,
according to
\begin{equation} \label{eqn:BasicStateEvolution}
X_n = A X_{n-1} + B_U U_n + B_S S_n    
\text{~~~and~~~}
Y_n = C X_n
\end{equation}
where $n=1,2,\ldots,$ indexes the pixels in the given row (or column),
and
\begin{equation}
A = \mleft(\begin{array}{cc} 
      I & I \\
      0 & I
    \end{array}\mright),\;
B_U = \mleft(\begin{array}{c} 
      I \\
      0
    \end{array}\mright),\;
B_S = \mleft(\begin{array}{c} 
      0 \\
      I
    \end{array}\mright),\;
C = (I, 0),
\end{equation}
where $I$ is a $3\times 3$ identity matrix for color images and \mbox{$I=1$} for grayscale images.
Note that the latent pixel value $Y_n$ is simply the upper part of $X_n$. 
Note also that, without an input (i.e., for $U_n=S_n=0$ for all $n$), 
the pixel values $Y_1, Y_2, \ldots$
follow a straight line in the intensity space, 
with a slope defined by the (constant) lower part of $X_n$.

The inputs $U_n$ and $S_n$ in (\ref{eqn:BasicStateEvolution})
will be referred to as level step inputs and slope noise inputs, respectively.
The level step inputs $U_n$ are intended to be sparse (as detailed below),
with a nonzero value of $U_n$ indicating an edge in the image.
In smooth parts of the image, $U_n$ is intended to be (approximately) zero,
but the slope noise $S_n$ can change the slope of the line model (\ref{eqn:BasicStateEvolution}),
as will be discussed in Section~\ref{sec:SlopeNoise}.


The state space models for the different rows and columns are independent,
but coupled by the condition that the latent pixel value $Y_n$
in the pertinent row model equals its value in the pertinent column model,
as illustrated in \Fig{fig:BasicModel}.

The inputs $U_n$ are i.i.d.\ random with a sparsifying prior
\begin{equation} \label{eqn:pUn}
p(u_n) \propto \exp\mleft( -\beta \| u_n \|^p \mright)
\end{equation}
with $0<p \leq 2$. 
(A fixed value $p\approx 0.3$ seems to work well.)

For computational tractability, 
we will use the well-known 
NUP representation of (\ref{eqn:pUn}) as
\begin{equation} \label{eqn:priorUn}
\exp\mleft( -\beta \| u_n \|^p \mright)
= 
\max_{\sigma_{U_n}^2} g(\sigma_{U_n}) \exp\mleft( -\frac{\| u_n \|^2}{2 \sigma_{U_n}^2} \mright)
\end{equation}
with a suitable function $g$.
The maximizing $\sigma_{U_n}^2$ in (\ref{eqn:priorUn}) is
well-known \cite{loeliger2023nup, bach2012optimization} to be 
\begin{equation} \label{eqn:updateSigmaUn}
\sigma_{U_n}^2 = \frac{\| u_n \|^{2-p}}{\beta p}
\end{equation}

\subsection{Slope Noise $S_n$ with Piecewise Constant Variance}
\label{sec:SlopeNoise}

The slope noise inputs $S_n$ are independent zero-mean Gaussian 
with variance $R_n^{-2}$ (or covariance matrix $R_n^{-2}I$).
The (scalar) parameters $R_n$ are not fixed,
but modeled as (approximately) piecewise constant. 
Specifically, we use the simple state space model
\begin{equation} \label{eqn:ScalarVarianceSSM}
R_n = R_{n-1} + \Delta_n,
\end{equation}
with i.i.d.\ random inputs $\Delta_n$ 
with a sparsifying prior
\begin{equation}\label{eqn:deltan}
p(\Delta_n) \propto \exp\mleft( -\beta_\Delta \| \Delta_n \|^p \mright),
\end{equation}
for which we use the same NUP representation as in (\ref{eqn:priorUn}), i.e.,
\begin{equation} \label{eqn:updateSigmaDeltan}
\sigma_{\Delta_n}^2 = \frac{\| \Delta_n \|^{2-p}}{p \beta_\Delta}
\end{equation}

\begin{figure*}
\centering
\begin{tikzpicture}[x=1mm, y=1mm, >=latex]
\tikzset{%
opbox/.style = {draw, rectangle, inner sep=0mm, minimum width=5mm, minimum height=5mm},
stdbox/.style = {draw, rectangle, minimum width=7mm, minimum height=7mm},
blobbox/.style = {draw, fill=black, rectangle, inner sep=0mm, minimum width=1.75mm, minimum height=1.75mm},
empty/.style={circle, minimum size=0mm, inner sep=0mm},
}

\draw node[empty](origin) {};
\draw (origin)+(-5,0) node {$\cdots$};
\draw (origin)+(17.5,0) node[stdbox](matA) {$A$};
\draw[->] (origin) --  node[above,pos=0.4]{$X_{n-1}$} (matA);
%
\draw (matA)+(17.5,0) node[opbox](add_level) {$+$};
\draw[->] (matA) -- (add_level);
\draw (add_level)+(20,0) node[opbox](add_slopechange) {$+$};
\draw[->] (add_level) -- (add_slopechange);
\draw (add_slopechange)+(17.5,0) node[opbox](equX) {$=$};
\draw[->] (add_slopechange) -- (equX);
\draw (equX)+(17.5,0) node[empty] (end) {};
\draw (equX) -- node[above,pos=0.5]{$X_n$} (end);
\draw ($(equX)!0.6!(end)$) node[anchor=north] { \annotdim{2(6)} };
\draw (end)+(5,0) node {$\cdots$};

\draw (add_level)+(0,12.5) node[stdbox](Blevel) {$B_U$};
\draw[->] (Blevel)--(add_level);
\draw (Blevel)+(0,17.5) node[opbox,label={above:$\calN(0,\sigma_{U_n}^2 I)$}](levelnoise) {};
\draw[->] (levelnoise)-- node[left]{$U_n$} (Blevel);
\draw ($(levelnoise)!0.5!(Blevel)$) node[anchor=west] { \annotdim{1(3)} };
\draw (levelnoise)+(-17.5,0) node[opbox,label={west:$g$}](gU) {};
\draw[->] (gU) -- node[below,pos=0.5]{$\sigma_{U_n}$} (levelnoise);

\draw (add_slopechange)+(0,12.5) node[stdbox](Bslope) {$B_S$};
\draw[->] (Bslope)--(add_slopechange);
\draw (Bslope)+(0,18.75) node[opbox,label={right:$e^{-r_n^2 \| s_n \|^2/2}$}](slopenoise) {};
\draw[->] (slopenoise) -- node[left,pos=0.625]{$S_n$} (Bslope);
\draw ($(slopenoise)!0.575!(Bslope)$) node[anchor=west] { \annotdim{1(3)} };
\draw (slopenoise)+(0,12.5) node[opbox](equSn) {$=$};
\draw[->] (equSn) -- (slopenoise);
\draw (equSn)+(12.5,0) node[opbox,label={above:$h$}](parSn) {};  
\draw[->] (parSn) -- (equSn);
\draw (parSn) -- node[above,pos=0.6]{$\theta_n$} +(15,0);  
\draw (slopenoise)+(-6,-6.25) 
     node[draw, rectangle, dashed, anchor=south west, minimum width=40mm, minimum height=27.5mm] (MLSP) {};
\draw (MLSP.east)+(0,0) node[right] {$\calN(0,r_n^{-2}I)$};

\draw (equSn)+(0,20) node[opbox](equSlopeSSM) {$=$};
\draw[->] (equSlopeSSM) -- node[left,pos=0.3]{$R_n''$} (equSn);
\draw (equSlopeSSM)+(17.5,0) node[empty] (endSlopeSSM) {};
\draw (equSlopeSSM) -- node[above,pos=0.5]{$R_n$} (endSlopeSSM);
\draw ($(equSlopeSSM)!0.6!(endSlopeSSM)$) node[anchor=north] { \annotdim{1(1)} };
\draw (endSlopeSSM)+(5,0) node {$\cdots$};
\draw (equSlopeSSM)+(-20,0) node[opbox](addSlopeSSM) {$+$};
\draw[->] (addSlopeSSM) -- node[above]{$R_n'$} (equSlopeSSM);
\draw (addSlopeSSM)+(-20,0) node(beginSlopeSSM) {};
\draw[->] (beginSlopeSSM) -- node[above,pos=0.4]{$R_{n-1}$} (addSlopeSSM);
\draw (beginSlopeSSM)+(-5,0) node {$\cdots$};
\draw (addSlopeSSM)+(0,15) node[opbox,label={above:$\calN(0,\sigma_{\Delta_n}^2)$}](noiseSlopeSMM) {};
\draw[->] (noiseSlopeSMM) -- node[right]{$\Delta_n$} (addSlopeSSM);
\draw (noiseSlopeSMM)+(-17.5,0) node[opbox,label={left:$g_R$}](gSlopeSSM) {};
\draw[->] (gSlopeSSM) -- node[below]{$\sigma_{\Delta_n}$} (noiseSlopeSMM);

\draw (equX)+(0,-12.5) node[stdbox](matC) {$C$};
\draw[->] (equX) -- (matC);
\draw (matC)+(0,-12.5) node[opbox](equPix) {$=$};
\draw[->] (matC) -- (equPix);
\draw (equPix)+(0,-15) node[opbox](add_obsnoise) {$+$};
\draw[->] (equPix) -- node[left,pos=0.5]{$Y_n$} (add_obsnoise);
\draw ($(equPix)!0.5!(add_obsnoise)$) node[anchor=west] { \annotdim{1(3)} };
\draw (add_obsnoise)+(0,-10) node[blobbox,label={east:$\breve y_n$}](obs) {};
\draw[->] (add_obsnoise) -- (obs);
\draw (add_obsnoise)+(-17.5,0) node[opbox,label={west:$\calN(0,\sigma_Z^2)$}](obsnoise) {};
\draw[->] (obsnoise) -- node[below]{$Z_n$} (add_obsnoise);

\draw (equPix)+(25,0) node[stdbox](matC_colum) {$C$};
\draw[->] (matC_colum) -- (equPix);
\draw (matC_colum)+(20,10) 
     node[draw, rectangle, minimum width=70mm, minimum height=10mm, rotate=90] 
     (column_model) { Column Model };
\draw[->] (column_model.north|-matC_colum) -- (matC_colum);

\end{tikzpicture}
\caption{\label{fig:BasicModel}%
Factor graph of the basic model, 
focussing on pixel~$n$ of some row
with noisy observation $\breve Y_n = \breve y_n$.
The small gray numbers (e.g., $\annotdim{1(3)}$) 
indicate the dimension of the corresponding variables for gray scale and color.
The dashed box uses the trick from \cite{loeliger2023nup}
to reduce the  MAP estimation of the slope noise scale factors $R_n$ to Gaussian message passing.
}
\end{figure*}
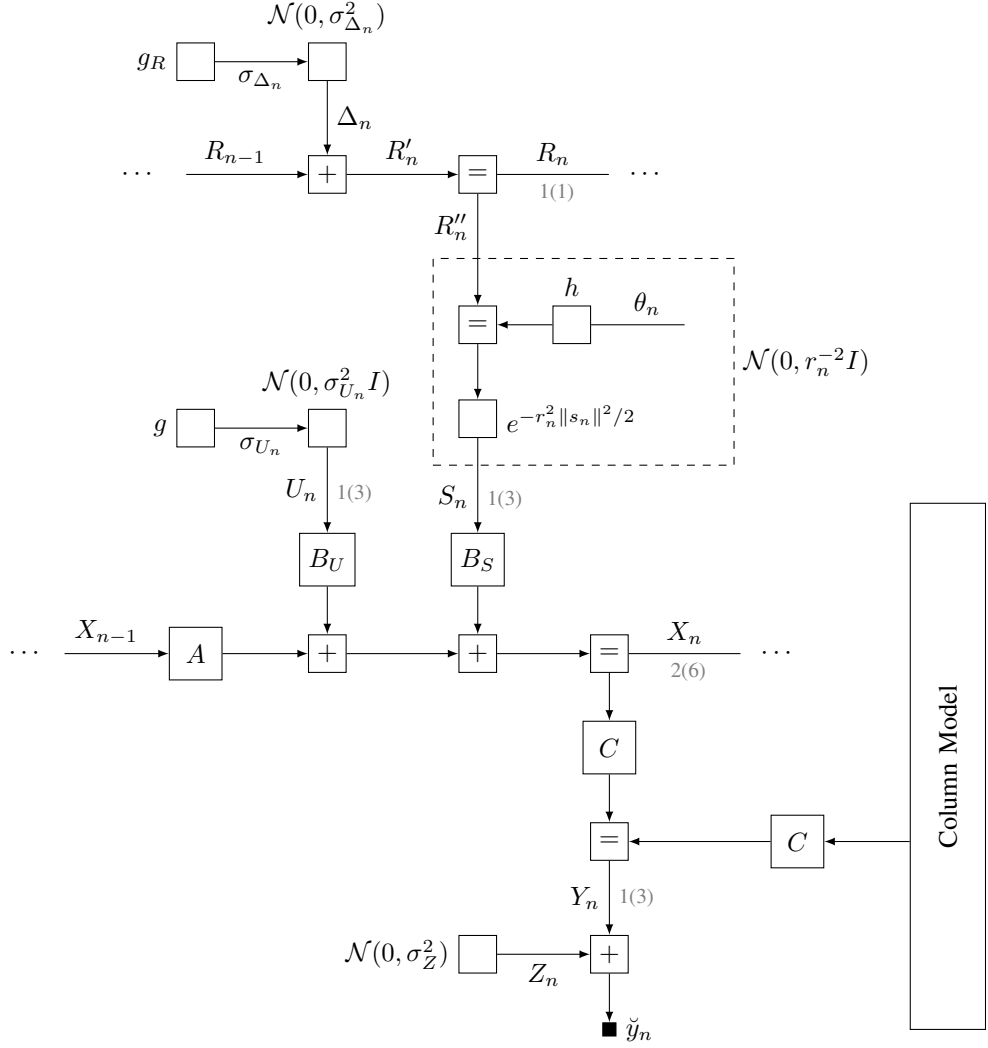

A factor graph of the entire basic model 
is shown in \Fig{fig:BasicModel}.
For computational tractability, 
the dashed box in \Fig{fig:BasicModel}
uses the NUP representation
\begin{equation} \label{eqn:MLSP:NUP}
|r_n|^m \exp\mleft( -r_n^2 \| s_n \|^2 /2 \mright) 
= \max_{\theta_n} h(r_n, \theta_n) \exp\mleft( -r_n^2 \| s_n \|^2 /2 \mright)
\end{equation}
proposed in \cite{loeliger2023nup}
(with $m=1$ for grayscale and $m=3$ for color),
with a suitable function $h$ and parameters $\theta_n$.
Using (\ref{eqn:MLSP:NUP}), \cite{loeliger2023nup} obtains 
a Gaussian message $\msgb{\mu}{R_n''}(r_n)$ 
with mean
\begin{equation} \label{eqn:MeanMLSP}
\msgb{m}{R''} = \beta_n \msgb{\sigma}{R''}^2
\end{equation}
and variance
\begin{equation} \label{eqn:Sigma2MLSP}
\msgb{\sigma}{R''}^2 = \mleft( \| s_n \|^2 - \hat r_n^{-2} + \beta_n |\hat r_n|^{-1} \mright)^{-1},
\end{equation}
where $\beta_n$ is a free parameter and 
$\hat r_n$ is the previous value of $R_n$ in the iterative estimation algorithm.
In a recent improvement 
due to Lukaj \cite{lukaj2026draft},
(\ref{eqn:MeanMLSP}) and (\ref{eqn:Sigma2MLSP}) are replaced by
\begin{equation} \label{eqn:MeanMLSP:new}
\msgb{m}{R''} = |\hat r_n|^{-1} \| s_n \|^{-2} 
\end{equation}
and variance
\begin{equation} \label{eqn:Sigma2MLSP:new}
\msgb{\sigma}{R''}^2 = \| s_n \|^{-2},
\end{equation}
respectively, 
which may formally be obtained by specializing 
(\ref{eqn:MeanMLSP}) and (\ref{eqn:Sigma2MLSP})
to $\beta_n = |\hat r_n|^{-1}$.

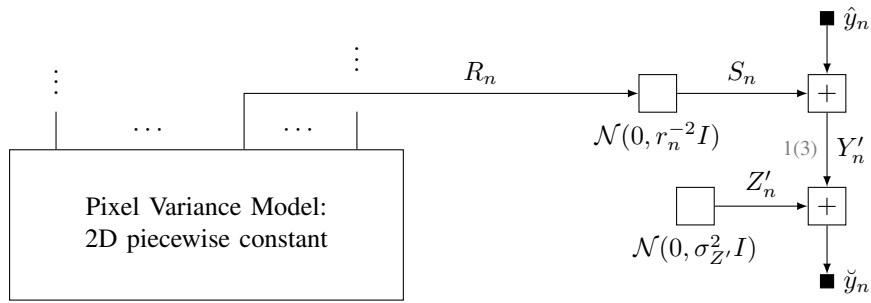
\begin{figure*}
\centering
\begin{tikzpicture}[x=1mm, y=1mm, >=latex]
\tikzset{%
opbox/.style = {draw, rectangle, inner sep=0mm, minimum width=5mm, minimum height=5mm},
stdbox/.style = {draw, rectangle, minimum width=7.5mm, minimum height=7.5mm},
blobbox/.style = {draw, fill=black, rectangle, inner sep=0mm, minimum width=1.75mm, minimum height=1.75mm},
empty/.style={circle, minimum size=0mm, inner sep=0mm},
}

\draw node[opbox,label={below: $\calN(0, r_n^{-2} I)$}] (norm1) {};
\draw (norm1)+(22.5,0) node[opbox] (add1) {$+$};
\draw[->] (norm1) -- node[above]{$S_n$} (add1);

\draw (add1)+(0,10) node[blobbox,label={right: $\hat y_n$}] (estimated) {};
\draw[->] (estimated) -- (add1);
\draw (add1)+(0,-15) node[opbox] (add2) {$+$};
\draw[->] (add1) -- node[right]{$Y'_n$} (add2);
\draw ($(add1)!0.5!(add2)$) node[anchor=east] { \annotdim{1(3)} };
\draw (add2)+(-17.5,0) node[opbox,label={below: $\calN(0,\sigma_{Z'}^2 I)$}] (norm2) {};
\draw[->] (norm2) -- node[above]{$Z_n'$} (add2);

\draw (add2)+(0,-10) node[blobbox,label={right: $\breve y_n$}] (obs) {};
\draw[->] (add2) -- (obs);


\draw (norm1)+(-60,-17.5) node[draw, rectangle, minimum width=50mm, minimum height=20mm] 
      (2Dmodel) { \parbox{50mm}{\centering Pixel Variance Model:\\ 2D piecewise constant} };
\draw[->] (2Dmodel.north)+(5,0) |- node[above,pos=0.8]{$R_n$} (norm1);
\draw (2Dmodel.north)++(-20,0) -- +(0,5);
\draw (2Dmodel.north)++(-20,10) node {$\vdots$};
\draw (2Dmodel.north)++(-7.5,2.5) node {$\cdots$};
\draw (2Dmodel.north)++(12.5,2.5) node {$\cdots$};
\draw (2Dmodel.north)++(20,0) -- +(0,5);
\draw (2Dmodel.north)++(20,13) node {$\vdots$};

\end{tikzpicture}
\vspace{1ex}
\caption{\label{fig:Phase2Model}%
Factor graph of 
the augmentation of Section~\ref{sec:Phase2},
focussing on a single pixel
with noisy observation $\breve Y_n = \breve y_n$.
(The variables $R_n$ and $S_n$ are unrelated 
 to the variables with these names in \Fig{fig:BasicModel}.)
}
\end{figure*}

\begin{figure*}
\begin{center}
\begin{minipage}{0.19\linewidth}
\centering
\begin{tikzpicture}[spy using outlines={rectangle, magnification=3, size= 42.5pt}]
\node[anchor=south west, inner sep=0] at (0,0){\includegraphics[height=90pt]{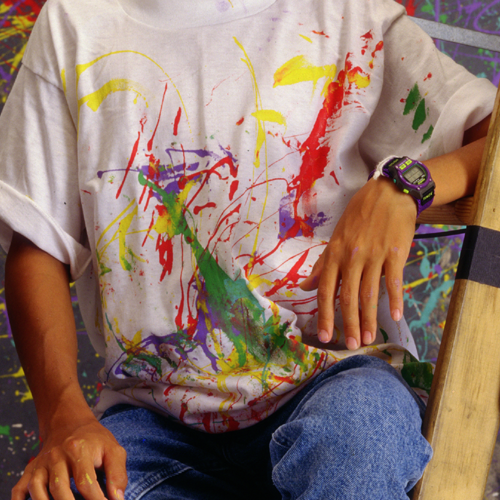}};
\spy [blue] on (0.8,2.3) in node [right] at (0.02,-30pt);
\spy [red] on (1.2,1.4) in node [right] at (1.68,-30pt);
\end{tikzpicture}
\captionsetup{justification=centering}
\caption*{Ground Truth\\PSNR/LPIPS}
\end{minipage}
\begin{minipage}{0.19\linewidth}
\centering
\begin{tikzpicture}[spy using outlines={rectangle, magnification=3, size= 42.5pt}]
\node[anchor=south west, inner sep=0] at (0,0){\includegraphics[height=90pt]{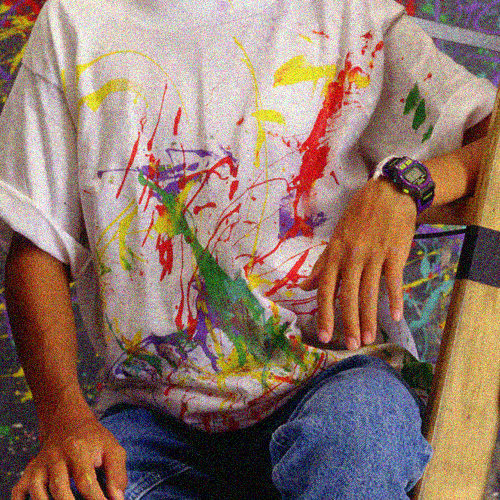}};
\spy [blue] on (0.8,2.3) in node [right] at (0.02,-30pt);
\spy [red] on (1.2,1.4) in node [right] at (1.68,-30pt);
\end{tikzpicture}
\captionsetup{justification=centering}
\caption*{Given (noisy)\\22.3/0.329}
\end{minipage}
\begin{minipage}{0.19\linewidth}
\centering
\begin{tikzpicture}[spy using outlines={rectangle, magnification=3, size= 42.5pt}]
\node[anchor=south west, inner sep=0] at (0,0){\includegraphics[height=90pt]{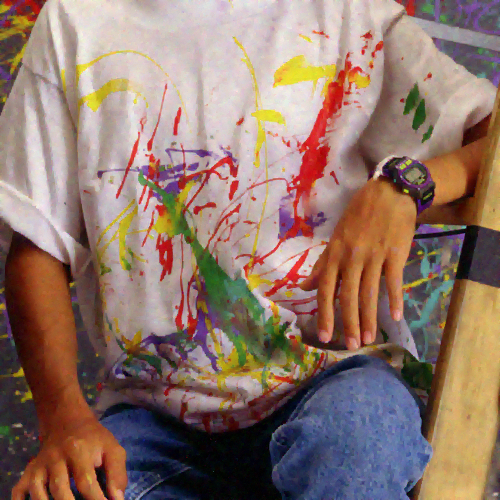}};
\spy [blue] on (0.8,2.3) in node [right] at (0.02,-30pt);
\spy [red] on (1.2,1.4) in node [right] at (1.68,-30pt);
\end{tikzpicture}
\captionsetup{justification=centering}
\caption*{\textbf{Proposed}\\\underline{30.0}/\textbf{0.082}}
\end{minipage}
\begin{minipage}{0.19\linewidth}
\centering
\begin{tikzpicture}[spy using outlines={rectangle, magnification=3, size= 42.5pt}]
\node[anchor=south west, inner sep=0] at (0,0){\includegraphics[height=90pt]{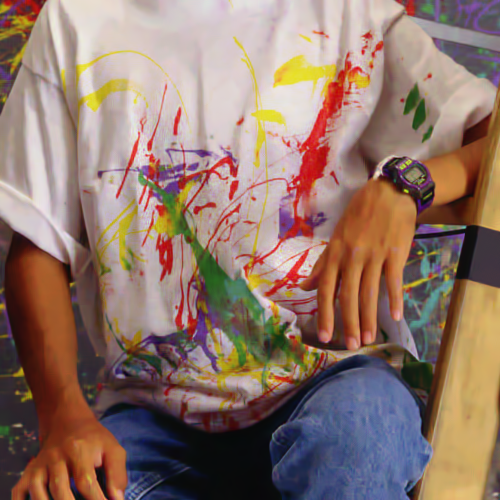}};
\spy [blue] on (0.8,2.3) in node [right] at (0.02,-30pt);
\spy [red] on (1.2,1.4) in node [right] at (1.68,-30pt);
\end{tikzpicture}
\captionsetup{justification=centering}
\caption*{BM3D\\\textbf{30.9}/0.139}
\end{minipage}
\begin{minipage}{0.19\linewidth}
\centering
\begin{tikzpicture}[spy using outlines={rectangle, magnification=3, size= 42.5pt}]
\node[anchor=south west, inner sep=0] at (0,0){\includegraphics[height=90pt]{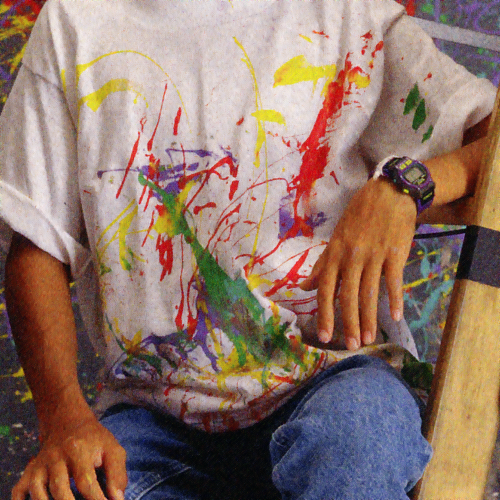}};
\spy [blue] on (0.8,2.3) in node [right] at (0.02,-30pt);
\spy [red] on (1.2,1.4) in node [right] at (1.68,-30pt);
\end{tikzpicture}
\captionsetup{justification=centering}
\caption*{ZS-N2N\\29.5/\underline{0.111}}
\end{minipage}
\caption{\label{fig:zoomin4}%
Example of denoising and comparison with two prior-art methods 
(BM3D \cite{dabov2006image, dabov2007image} and ZS-N2N \cite{mansour2023zero}).
The image is \#5 in McMaster18 \cite{zhang2011color}.
The (actual) added Gaussian noise level is $\sigma = 20/255$,
the assumed parameter $\sigma_Z$ is $1/21$.
\textbf{Bold} and \underline{underline} indicate the best and second best scores, respectively.
The proposed method produces arguably fewer artifacts.
}
\end{center}
\end{figure*}

\subsection{Estimation Algorithm}
\label{sec:BasicAlgorithm}

The model is constructed such that,
for fixed $\sigma_{U_n}^2$ and $R_n=r_n$ (for all $n$ in all rows and all columns),
the variables $U_n, S_n, X_n, Y_n$ are jointly Gaussian,
and computing the joint MAP estimates of these variables 
amounts to a sparse least squares problem
(where rows and columns remain coupled).
In our experience so far, 
the most attractive method 
to solve this least squares problem is the conjugate gradient algorithm.

Moreover, for fixed $s_n$ (for all $n$ in all rows and all columns),
the models for $R_n$ decouple into independent cycle-free factor graphs
for every row and every column. 
For fixed $\sigma_{\Delta_n}$ (for all $n$) 
and using (\ref{eqn:MeanMLSP:new}) and (\ref{eqn:Sigma2MLSP:new}),
these models for $R_n$ become linear Gaussian,
and their MAP estimates can be computed 
by 
forward-backward (or backward-forward) 
scalar Gaussian message passing.

We thus obtain the following cyclic optimization algorithm 
for computing the joint MAP estimate of all variables
by repeating the following steps until convergence:
\begin{enumerate}
\item
For fixed $\sigma_{U_n}^2$ and \mbox{$R_n=r_n$} (for all $n$ in all rows and columns),
compute the joint MAP estimate of $U_n, S_n, X_n, Y_n$
by solving the corresponding sparse least-squares problem with the conjugate gradient algorithm.
\item
For fixed $U_n = u_n$
, recompute $\sigma^2_{U_n}$ using (\ref{eqn:updateSigmaUn}).
\item
For fixed $\sigma_{\Delta_n}^2$, fixed \mbox{$S_n=s_n$}
, and using (\ref{eqn:MeanMLSP:new}) and (\ref{eqn:Sigma2MLSP:new}),
recompute all $R_n$ 
by scalar Gaussian message passing.
\item
For fixed $R_n=r_n$
, recompute $\sigma^2_{\Delta_n}$ 
by (\ref{eqn:updateSigmaDeltan}).
\end{enumerate}

It is easily seen that the joint MAP estimate of all these variables
is a fixed point of this algorithm;
beyond that, the algorithm has no guarantees, but empirically works well.
Five iterations 
usually suffice.
The computational complexity is dominated by the conjugate gradient algorithm.

\subsection{Augmented Model and Algorithm}
\label{sec:Phase2}

The (optional) augmentation of the model proposed in this section
aims at getting back some details (e.g., texture) into the smooth parts 
that might have been oversmoothed by the basic model.
To this end, 
the latent pixels $Y_n$ are reinterpreted as a local background,
which is estimated with the algorithm of Section~\ref{sec:BasicAlgorithm}.
The final estimates of the pixels are obtained 
by locally adaptive L2 regularization against this background.

Specifically,
let $\hat y_n$ be the value of $Y_n$ that was estimated as described above.
The actual pixel value $Y'_n$ is now modeled as
\begin{equation}
Y_n' = \hat y_n + S_n,
\end{equation}
where $S_n$ is the deviation%
\footnote{In this section, we reuse several variable names 
(including $S_n$ and $R_n$)
from Sections \ref{sec:BasicModel} and~\ref{sec:SlopeNoise}  
for new variables, which facilitates the discussion.} 
from the local background.
As in (\ref{eqn:ObsYn}),
we observe
\begin{equation}
\breve Y_n = Y_n' + Z'_n,
\end{equation}
where $Z'_n$ is white Gaussian noise with fixed variance $\sigma_{Z'}^2$,
with $\sigma_{Z'}^2 = \sigma_Z^2$ being a viable choice.

The local deviations $S_n$ are also modeled 
as independent zero-mean Gaussian random variables, 
but with individual variances $R_n^{-2}$.

Specifically, $R_n$ are modeled as (approximately) piecewise constant, 
as is summarily illustrated in \Fig{fig:Phase2Model}.
This is implemented by scalar state space models 
with sparse increments $\Delta_n$ as in (\ref{eqn:ScalarVarianceSSM}), 
one such model for each row and each column;
these row models and column models are independent,
but coupled by the condition 
that the value of $R_n$
in the pertinent row model equals its value in the pertinent column model.

The node ``$\calN(0, r_n^{-2} I)$''
in \Fig{fig:Phase2Model}
is implemented like the dashed box in \Fig{fig:BasicModel}.

We then obtain
the following cyclic optimization algorithm 
for computing the joint MAP estimate of all variables 
by repeating the following steps until convergence:
\begin{enumerate}
\item
For fixed $R_n = r_n$, compute the (trivial) MAP estimate of $Y_n'$ for all $n$. 
\item
For fixed $Y_n' = y_n'$, compute the Gaussian backward message through $R_n$ 
using (\ref{eqn:MeanMLSP:new}) and~(\ref{eqn:Sigma2MLSP:new}).
\item
For fixed variances $\sigma_{\Delta_n}^2$ and fixed backward messages through $R_n$,
compute the joint MAP estimate of all $R_n$
using the conjugate gradient algorithm.
\item
For fixed $R_n=r_n$, 
update the variances $\sigma_{\Delta_n}^2$
using (\ref{eqn:updateSigmaDeltan}).
\end{enumerate}
As in Section~\ref{sec:BasicAlgorithm},
the joint MAP estimate of all variables
is a fixed point of this algorithm,
and the computational complexity is dominated by the conjugate gradient algorithm.

\begin{figure*}
\centering
\begin{minipage}{0.19\linewidth}
\centering
\includegraphics[height = 90pt]{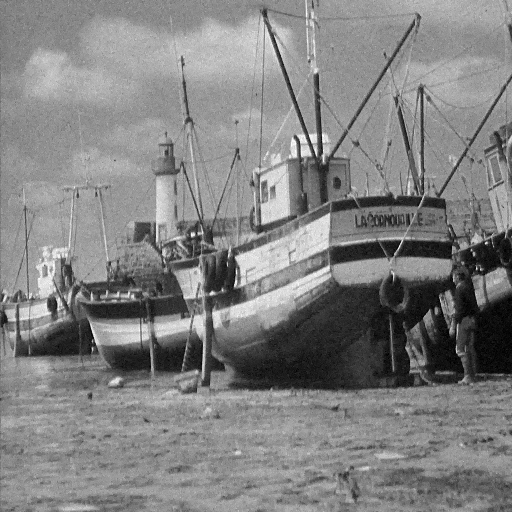}
\captionsetup{justification=centering}
\caption*{$\sigma_Z^{-1}$ = 40\\[0.5ex]
30.3/0.201}
\end{minipage}
\begin{minipage}{0.19\linewidth}
\centering
\includegraphics[height = 90pt]{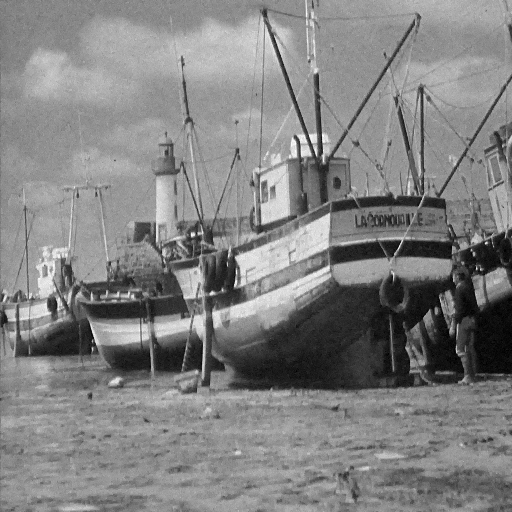}
\captionsetup{justification=centering}
\caption*{$\sigma_Z^{-1}$ = 35\\[0.5ex]
32.1/0.122}
\end{minipage}
\begin{minipage}{0.19\linewidth}
\centering
\includegraphics[height = 90pt]{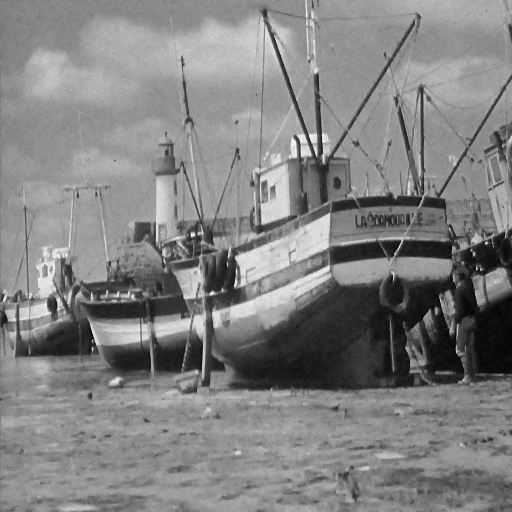}
\captionsetup{justification=centering}
\caption*{$\sigma_Z^{-1}$ = 30\\[0.5ex]
\textbf{32.6}/\textbf{0.096}}
\end{minipage}
\begin{minipage}{0.19\linewidth}
\centering
\includegraphics[height = 90pt]{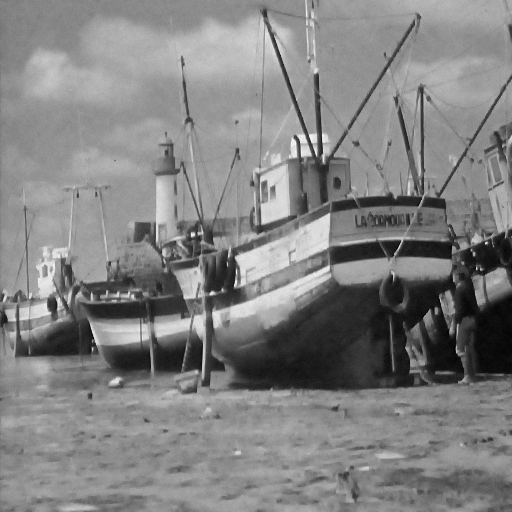}
\captionsetup{justification=centering}
\caption*{$\sigma_Z^{-1}$ = 25\\[0.5ex]
32.1/0.114}
\end{minipage}
\begin{minipage}{0.19\linewidth}
\centering
\includegraphics[height =90pt]{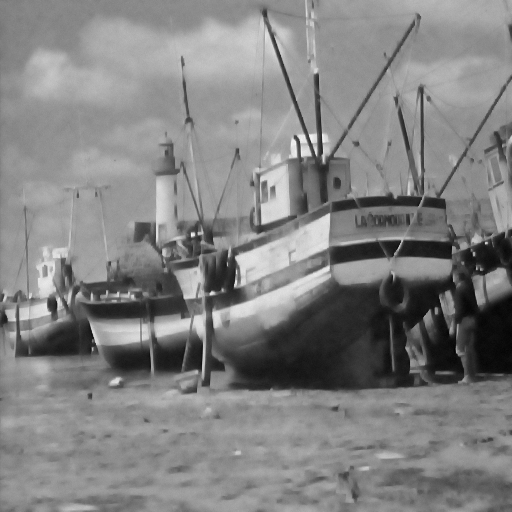}
\captionsetup{justification=centering}
\caption*{$\sigma_Z^{-1}$ = 20\\[0.5ex]
31.1/0.152}
\end{minipage}
\caption{\label{fig:boat}%
Effect of the parameter $\sigma_Z$, with best numerical scores 
(PSNR and LPIPS) in \textbf{bold}.
The image is Boat in \cite{dabov2006image}, 
the (actual) Gaussian noise level is $\sigma = 10/255$.
}
\end{figure*}

\begin{table*}[t]
\begin{center}
\begin{tabular}{@{}clc@{~~~}cc@{}c@{}cc@{}c@{}cc@{}c@{}cc@{}}
\toprule
&&& \multicolumn{5}{c}{McMaster18 (color)} & \hspace{2em} & \multicolumn{5}{c}{Set12 (grayscale)}
\\
\cmidrule{4-8}\cmidrule{10-14}
\multirow{2}{*}{\rotatebox{90}{Local}}
&&&  \multicolumn{2}{c}{$\sigma = 10/255$} & \hspace{1.5em} & \multicolumn{2}{c}{$\sigma = 20/255$}
&& \multicolumn{2}{c}{$\sigma = 10/255$} & \hspace{1.5em} & \multicolumn{2}{c}{$\sigma = 20/255$}
\\
\cmidrule{4-5}\cmidrule{7-8}
\cmidrule{10-11}\cmidrule{13-14}
&&&   PSNR  &   LPIPS   &&  PSNR  &   LPIPS
&&   PSNR  &   LPIPS   &&  PSNR  &   LPIPS
\\
\midrule
& Noisy image
&&28.5    &0.149    &&22.7    & 0.365
&&28.1    &0.251    &&22.2    & 0.513
\\
\midrule
\checkmark & TV \cite{rudin1992nonlinear, chambolle2004algorithm}
&&32.8    &0.092    &&29.4    & 0.165
&&31.8    &0.105    &&28.8    & 0.172
\\
\checkmark & SNUV \cite{ma2022smoothed}
&&33.0    &\underline{0.049}    &&29.9    & \underline{0.116} 
&&32.4    &\underline{0.080}    &&28.8    &  0.171
\\
\checkmark & Proposed (augm.)
&&33.4    &\textbf{0.043}    &&30.4    & \textbf{0.114} 
&&\underline{32.7}    &\textbf{0.075}    &&\underline{29.2}    & \underline{0.158}  
\\
\checkmark & Proposed (basic)
&&\underline{34.1}    & 0.052   &&\underline{30.6}    & 0.123 
&&32.6    &0.096    &&29.1    &0.181 
\\
\midrule
-- & BM3D \cite{dabov2006image, dabov2007image} 
&&\textbf{35.0}    &0.072    &&\textbf{31.3}    & 0.144 
&&\textbf{33.7}    &0.084    &&\textbf{30.6}    &  \textbf{0.143}
\\
-- & ZS-N2N \cite{mansour2023zero}
&&34.1    &0.054    &&30.2    &  0.122
&&32.3    &0.082    &&28.5    &  0.182
\\
\bottomrule
\end{tabular}
\end{center}
\caption{\label{table:compare}%
Denoising with the proposed method 
(with and without the augmentation of Section~\ref{sec:Phase2})
and several prior-art methods.
The table shows the average numerical scores 
for data sets McMaster18 \cite{zhang2011color} 
and Set12 \cite{zhang2017beyond, zhang2018ffdnet},
for two different levels of added Gaussian noise.
\textbf{Bold} and \underline{underline} indicate the best and second best scores, respectively.}
\end{table*}

\section{Applications and Experiments}
\label{sec:Experiments}

\subsection{Denoising of Gaussian Noise}

A first example is shown in \Fig{fig:zoomin4}. 
In this example, and in all examples of this subsection,
the given noisy image is obtained from a known ground truth
by adding i.i.d.\ Gaussian noise.
The numerical scores are the standard peak signal-to-noise ratio (PSNR)
and the perceptual loss (LPIPS) from \cite{zhang2018unreasonable}.

\begin{figure}[t]
\hfill
\begin{minipage}{0.3\linewidth}
\centering
\includegraphics[height = 75pt,
trim=5cm 8cm 11cm 8cm,
clip
]{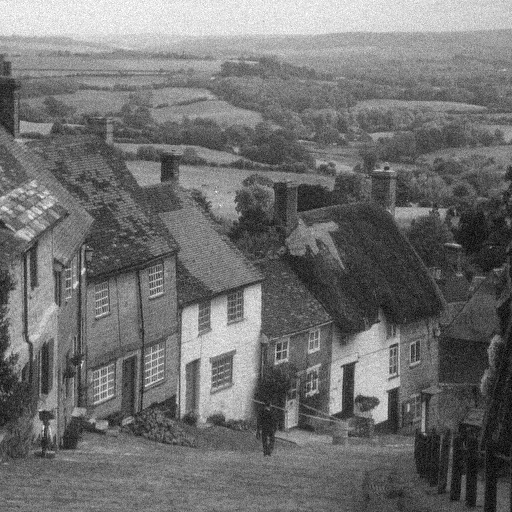}
\captionsetup{justification=centering}
\caption*{Noisy}
\end{minipage}
\hfill
\begin{minipage}{0.3\linewidth}
\centering
\includegraphics[height = 75pt,
trim=5cm 8cm 11cm 8cm,
clip
]{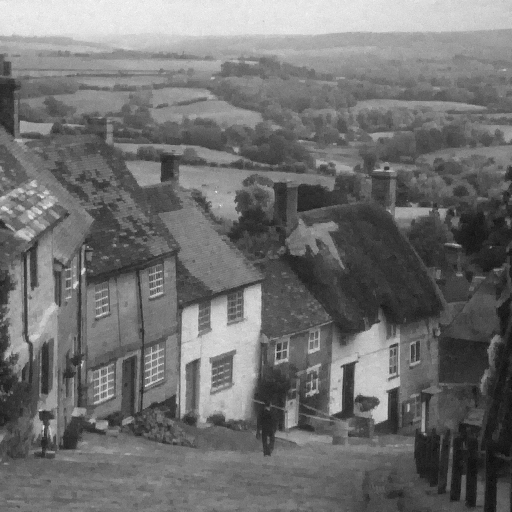}
\captionsetup{justification=centering}
\caption*{Proposed (augm.)}
\end{minipage}
\hfill
\begin{minipage}{0.3\linewidth}
\centering
\includegraphics[height = 75pt,
trim=5cm 8cm 11cm 8cm,
clip
]{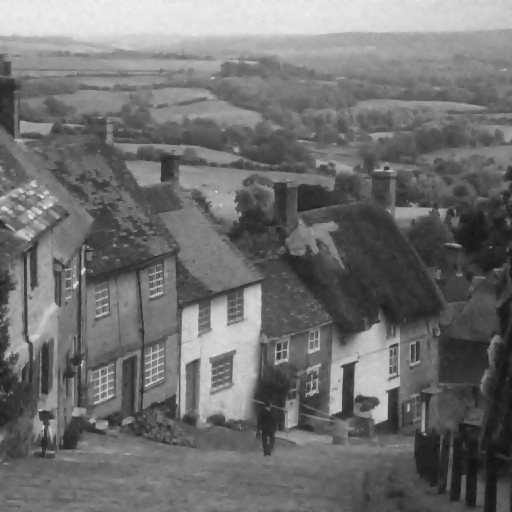}
\captionsetup{justification=centering}
\caption*{Proposed (basic)}
\end{minipage}
\hfill
\hfill
\caption{\label{fig:augm}%
Denoising with basic model and augmented model.
The image is Hill (cropped)
from \cite{dabov2006image} with (actual) Gaussian noise level $\sigma = 10/255$.
}
\end{figure}

The effect of the user-defined parameter $\sigma_Z^2$ 
is demonstrated 
in \Fig{fig:boat}.
The effect of the augmented model (vs.\ the basic model) is demonstrated
in \Fig{fig:augm}. 

Table~\ref{table:compare} gives a systematic comparison 
with several prior-art methods that do not need to be trained on datasets.
Note that 
BM3D and ZS-N2N 
use nonlocal information within the image
and are prone to produce nonlocal artifacts. 
In these experiments, the parameter $\sigma_Z$ of the proposed method
was adapted to the noise level, but not to individual images.

\subsection{Contrast Enhancement}
\label{sec:Contrast}

\begin{figure}[t]
\hfill
\begin{minipage}{0.3\linewidth}
\centering
\includegraphics[height = 75pt,
trim=1.5cm 1cm 1.5cm 1.5cm,
clip
]{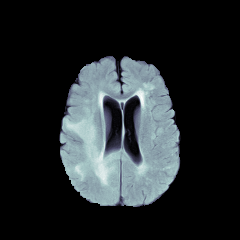}
\captionsetup{justification=centering}
\caption*{Given}
\end{minipage}
\hfill
\begin{minipage}{0.3\linewidth}
\centering
\includegraphics[height = 75pt,
trim=1.5cm 1cm 1.5cm 1.5cm,
clip
]{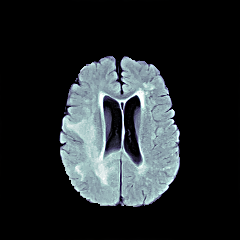}
\captionsetup{justification=centering}
\caption*{Enhanced}
\end{minipage}
\hfill
\hfill
\caption{\label{fig:brain}%
Example of contrast enhancement.
}
\end{figure}

Denoising and contrast enhancement are commonly considered 
as entirely different tasks requiring entirely different methods.
However, the proposed model can easily be used for contrast enhancement as follows.
First, the algorithm of Section~\ref{sec:BasicAlgorithm} is used to compute
estimates $\hat u_n$ of the level steps $U_n$. 
Second, the level steps are fixed to $U_n = \varphi(\hat u_n)$,
and all other variables are restimated. 
The function $\varphi(u_n)$ should be symmetric, concave for $u_n\geq 0$, 
and have slope $>1$ at $u_n=0$;
e.g., $\varphi(u_n) = \alpha \tanh(\beta u_n)$ with $\beta>1$.
An exemplary result with this method is shown in \Fig{fig:brain}.
The image in \Fig{fig:brain} is a slice of brain MRI scan from BraTS 2018 \cite{menze2014multimodal, bakas2017advancing, bakas2018identifying}.

\subsection{More Applications and Details}

More details, applications (including inpainting and non-Gaussian denoising), 
and examples are given in
the appendix of the arXiv version (arXiv:2601.08749).


\section{Conclusion}

We proposed a novel image model 
that is more flexible than prior explicit image models and
needs only one (interpretable) parameter to be set by the user.
The empirical performance for denoising 
is arguably at least as good as 
two specialized prior-art methods.
Moreover, the model and algorithms are easily adapted to other tasks 
including (but not limited to) contrast enhancement, non-Gaussian denoising, and inpainting.

Technically, and of independent interest,
the paper exemplifies model-based signal processing 
in the spirit of \cite{loeliger2007factor} 
with recent advances as summarized in \cite{loeliger2023nup}.

\newpage

\section{Appendix}

\subsection{Parameter Setting}
The default initializations for basic model are $\sigma_{U_{n}} = 1/10,
r_{n} = 500, \sigma_{\Delta_{n}} = 10$ for all $n$.
And the default initializations for augmented model are $ \sigma_{Z'} = \sigma_{Z}, r_{n} = 60, 
\sigma_{\Delta_{n}} = 10$ for all $n$.
For both basic and augmented model,
five iterations usually suffice.

\subsection{Basic Model and Augmented Model}
\begin{figure*}
\begin{center}
\begin{minipage}{0.24\linewidth}
\centering
\begin{tikzpicture}[spy using outlines={rectangle, magnification=6, size= 120pt}]
\node[anchor=south west, inner sep=0] at (0,0){\includegraphics[height=120pt]{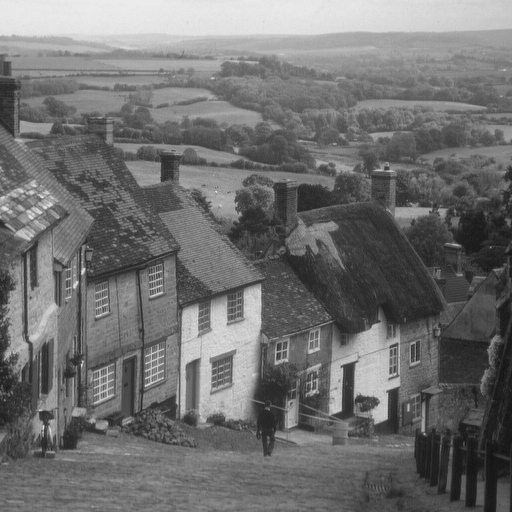}};
\spy [red] on (1.4, 2.1) in node [right] at (0.0,-70pt);
\end{tikzpicture}
\captionsetup{justification=centering}
\caption*{Ground Truth\\PSNR/LPIPS}
\end{minipage}
\begin{minipage}{0.24\linewidth}
\centering
\begin{tikzpicture}[spy using outlines={rectangle, magnification=6, size= 120pt}]
\node[anchor=south west, inner sep=0] at (0,0){\includegraphics[height=120pt]{figures/hill/hill_noisy.png}};
\spy [red] on (1.4, 2.1) in node [right] at (0.0,-70pt);
\end{tikzpicture}
\captionsetup{justification=centering}
\caption*{Noisy\\28.1/0.250}
\end{minipage}
\begin{minipage}{0.24\linewidth}
\centering
\begin{tikzpicture}[spy using outlines={rectangle, magnification=6, size= 120pt}]
\node[anchor=south west, inner sep=0] at (0,0){\includegraphics[height=120pt]{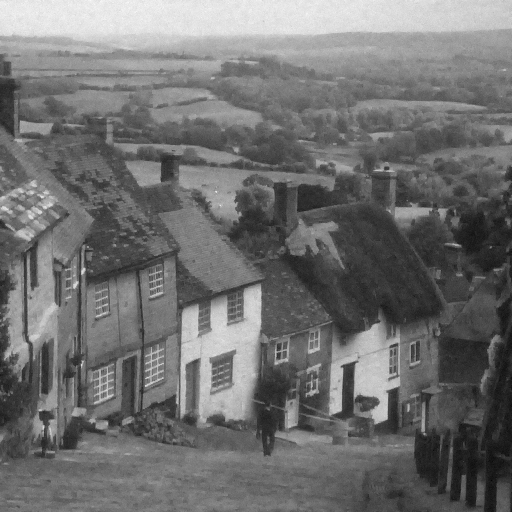}};
\spy [red] on (1.4, 2.1) in node [right] at (0.0,-70pt);
\end{tikzpicture}
\captionsetup{justification=centering}
\caption*{Proposed (augm.)\\\textbf{32.6}/\textbf{0.095}}
\end{minipage}
\begin{minipage}{0.24\linewidth}
\centering
\begin{tikzpicture}[spy using outlines={rectangle, magnification=6, size= 120pt}]
\node[anchor=south west, inner sep=0] at (0,0){\includegraphics[height=120pt]{figures/hill/hill_LA.png}};
\spy [red] on (1.4, 2.1) in node [right] at (0.0,-70pt);
\end{tikzpicture}
\captionsetup{justification=centering}
\caption*{Proposed (basic)\\32.3/0.162}
\end{minipage}
\caption{Visual comparison of image Hill in \cite{dabov2006image} for Gaussian noise level $\sigma = 10/255$.
The quantitative PSNR/LPIPS results are given below the image.
\textbf{Bold} indicates the best results.
}
\label{fig:zoominG}
\end{center}
\end{figure*}

The difference between basic model and augmented model is demonstrated in \Fig{fig:zoominG}.
The bricks in the reconstruction result of augmented model is much more obvious, indicating that
the augmented model indeed adds some details back when compared to the basic model.

\subsection{Effect of Parameter $\sigma_Z^2$}
\begin{figure*}
\centering
\begin{minipage}{0.19\linewidth}
\centering
\includegraphics[height = 90pt]{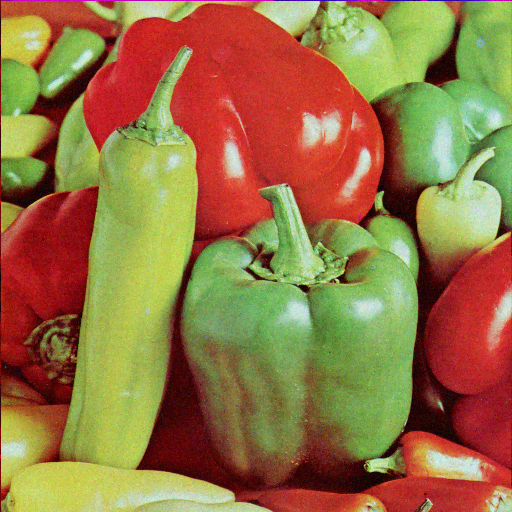}
\captionsetup{justification=centering}
\caption*{$\sigma_Z^{-1}$ = 40\\[0.5ex]
30.4/0.064}
\end{minipage}
\begin{minipage}{0.19\linewidth}
\centering
\includegraphics[height = 90pt]{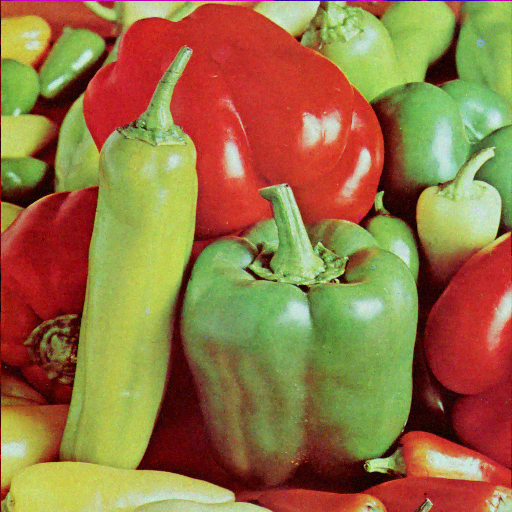}
\captionsetup{justification=centering}
\caption*{$\sigma_Z^{-1}$ = 35\\[0.5ex]
31.7/0.036}
\end{minipage}
\begin{minipage}{0.19\linewidth}
\centering
\includegraphics[height = 90pt]{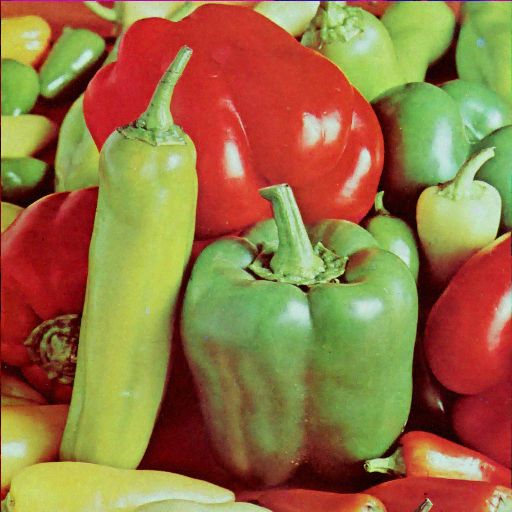}
\captionsetup{justification=centering}
\caption*{$\sigma_Z^{-1}$ = 30\\[0.5ex]
33.0/\textbf{0.029}}
\end{minipage}
\begin{minipage}{0.19\linewidth}
\centering
\includegraphics[height = 90pt]{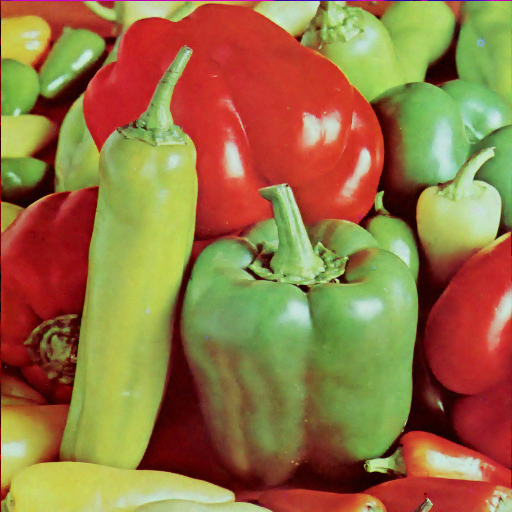}
\captionsetup{justification=centering}
\caption*{$\sigma_Z^{-1}$ = 25\\[0.5ex]
\textbf{33.5}/0.079}
\end{minipage}
\begin{minipage}{0.19\linewidth}
\centering
\includegraphics[height =90pt]{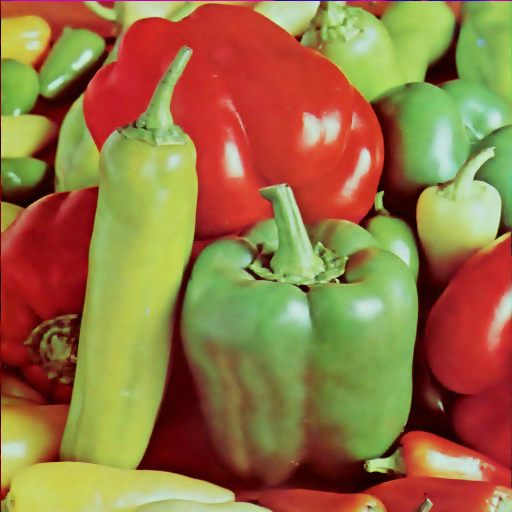}
\captionsetup{justification=centering}
\caption*{$\sigma_Z^{-1}$ = 20\\[0.5ex]
32.9/0.145}
\end{minipage}
\begin{minipage}{0.19\linewidth}
\vspace{2ex}
\centering
\includegraphics[height = 90pt]{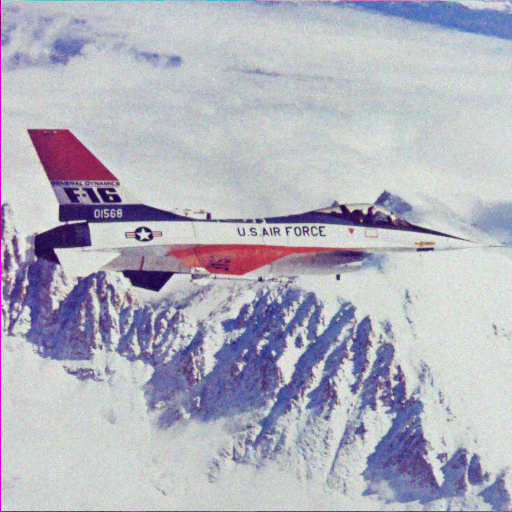}
\captionsetup{justification=centering}
\caption*{$\sigma_Z^{-1}$ = 40\\[0.5ex]
30.7/0.152}
\end{minipage}
\begin{minipage}{0.19\linewidth}
\vspace{2ex}
\centering
\includegraphics[height = 90pt]{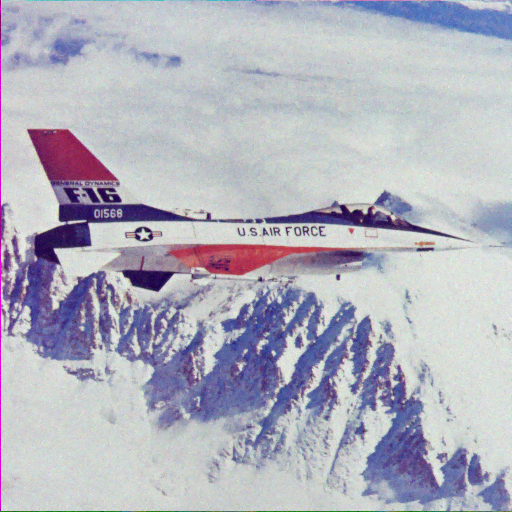}
\captionsetup{justification=centering}
\caption*{$\sigma_Z^{-1}$ = 35\\[0.5ex]
32.2/0.101}
\end{minipage}
\begin{minipage}{0.19\linewidth}
\vspace{2ex}
\centering
\includegraphics[height = 90pt]{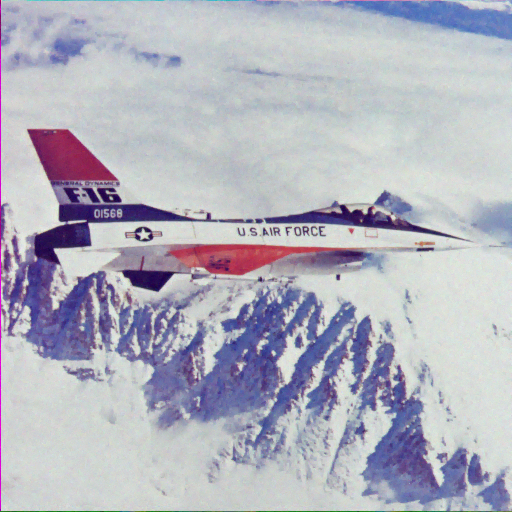}
\captionsetup{justification=centering}
\caption*{$\sigma_Z^{-1}$ = 30\\[0.5ex]
34.0/\textbf{0.066}}
\end{minipage}
\begin{minipage}{0.19\linewidth}
\vspace{2ex}
\centering
\includegraphics[height = 90pt]{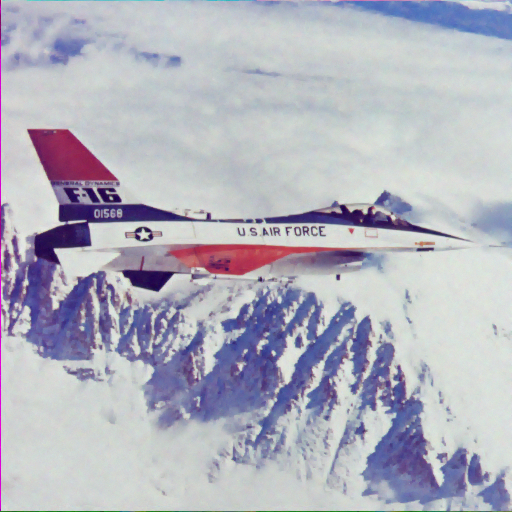}
\captionsetup{justification=centering}
\caption*{$\sigma_Z^{-1}$ = 25\\[0.5ex]
\textbf{34.9}/0.096}
\end{minipage}
\begin{minipage}{0.19\linewidth}
\vspace{2ex}
\centering
\includegraphics[height =90pt]{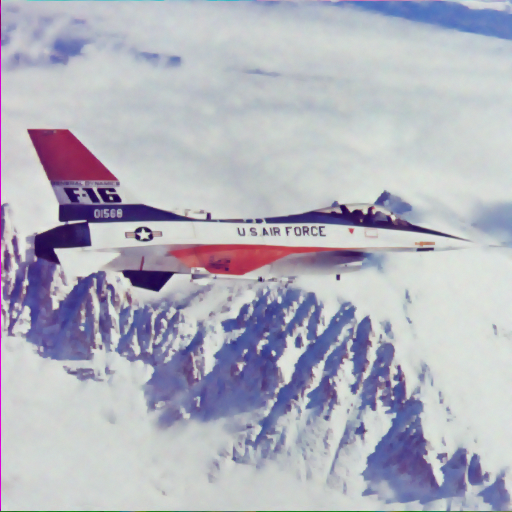}
\captionsetup{justification=centering}
\caption*{$\sigma_Z^{-1}$ = 20\\[0.5ex]
34.2/0.156}
\end{minipage}
\caption{
Effect of the parameter $\sigma_Z$, with best numerical scores (PSNR/LPIPS) in \textbf{bold}.
The images are Peppers and F16 in \cite{dabov2006image}, the (actual) Gaussian noise level is $\sigma = 10/255$.
}
\label{fig:pepper}
\end{figure*}

The effect of the user-defined parameter $\sigma_Z^2$ is demonstrated in other color image examples
in \Fig{fig:pepper}. With the increase of $\sigma_Z^2$, the denoised images are getting smoother.
It is noted that the best scores in BM3D and LPIPS may not be achieved by the same
parameter $\sigma _Z^2$.

\subsection{Poissonian-Gaussian Noise}
In the case of Poissonian-Gaussian noise, the observation $\breve{Y}_n$
is considered to be obtained by first applying Poisson corruption to pixel $X_n$
with peak intensity $\alpha$ (related to the gain of camera)
and then adding white Gaussian noise with variance $\sigma_Z^2$. Then we have
\begin{equation}
\breve{Y}_n = \alpha^{-1} \calP(\alpha X_n) + \calN(0,\sigma_Z^2).
\end{equation}
With the common approximation of Poisson noise as the Gaussian noise with equal mean and variance \cite{foi2008practical},
we get
\begin{equation}
\breve{Y}_n \approx  \alpha^{-1} \calN(\alpha X_n, \alpha X_n) + \calN(0,\sigma_Z^2).
\end{equation}
Then the noisy pixel $\breve{Y}_n$ can be regarded as the original pixel $X_n$ with an additive
white Gaussian noise whose variance is relative to the pixel intensity, 
\begin{equation}
\breve{Y}_n \approx  X_n  + \calN(0, \alpha^{-1} X_n + \sigma_Z^2).
\end{equation}
Therefore we introduce a new user-defined parameter $\alpha$ to model
the Poissonian-Gaussian noise.

The examples of denoising with Poissonian-Gaussian 
noise are demonstrated in Figure 8-10.
The actual Gaussian noise level is $\sigma = 10/255$ and actual Poissonian noise peak intensity is 100.
The assumed parameter $\sigma_Z$ is $1/35$ and $\alpha$ is 275.

\begin{figure*}
\begin{center}
\begin{minipage}{0.19\linewidth}
\centering
\begin{tikzpicture}[spy using outlines={rectangle, magnification=3, size= 42.5pt}]
\node[anchor=south west, inner sep=0] at (0,0){\includegraphics[height=90pt]{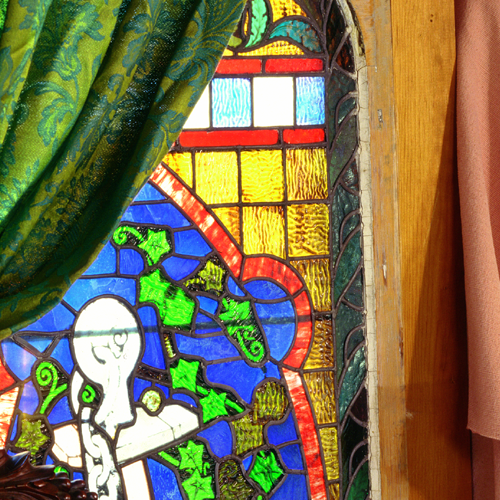}};
\spy [blue] on (0.8,2.3) in node [right] at (0.02,-30pt);
\spy [red] on (1.2,1.4) in node [right] at (1.68,-30pt);
\end{tikzpicture}
\captionsetup{justification=centering}
\caption*{Ground Truth\\PSNR/LPIPS}
\end{minipage}
\begin{minipage}{0.19\linewidth}
\centering
\begin{tikzpicture}[spy using outlines={rectangle, magnification=3, size= 42.5pt}]
\node[anchor=south west, inner sep=0] at (0,0){\includegraphics[height=90pt]{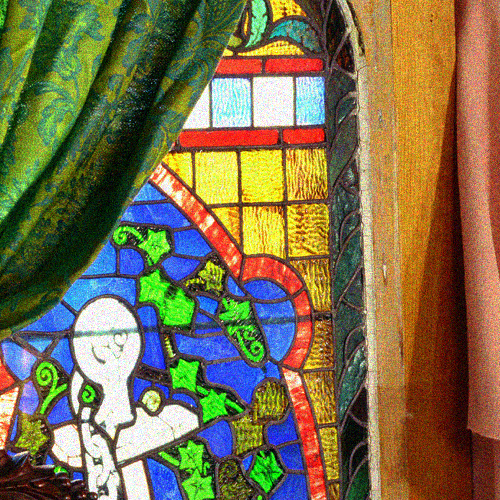}};
\spy [blue] on (0.8,2.3) in node [right] at (0.02,-30pt);
\spy [red] on (1.2,1.4) in node [right] at (1.68,-30pt);
\end{tikzpicture}
\captionsetup{justification=centering}
\caption*{Noisy\\23.1/0.151}
\end{minipage}
\begin{minipage}{0.19\linewidth}
\centering
\begin{tikzpicture}[spy using outlines={rectangle, magnification=3, size= 42.5pt}]
\node[anchor=south west, inner sep=0] at (0,0){\includegraphics[height=90pt]{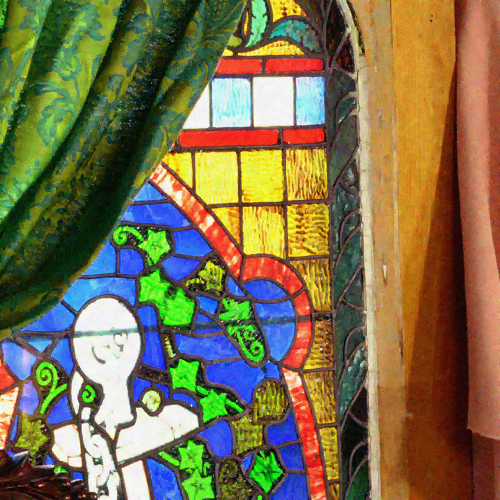}};
\spy [blue] on (0.8,2.3) in node [right] at (0.02,-30pt);
\spy [red] on (1.2,1.4) in node [right] at (1.68,-30pt);
\end{tikzpicture}
\captionsetup{justification=centering}
\caption*{\textbf{Proposed}\\27.5/\textbf{0.059}}
\end{minipage}
\begin{minipage}{0.19\linewidth}
\centering
\begin{tikzpicture}[spy using outlines={rectangle, magnification=3, size= 42.5pt}]
\node[anchor=south west, inner sep=0] at (0,0){\includegraphics[height=90pt]{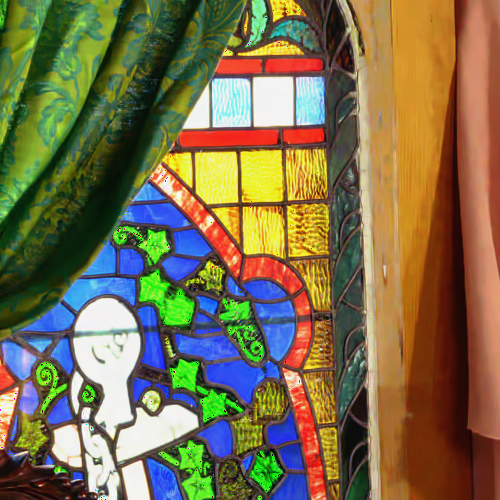}};
\spy [blue] on (0.8,2.3) in node [right] at (0.02,-30pt);
\spy [red] on (1.2,1.4) in node [right] at (1.68,-30pt);
\end{tikzpicture}
\captionsetup{justification=centering}
\caption*{BM3D\\27.3/0.135}
\end{minipage}
\begin{minipage}{0.19\linewidth}
\centering
\begin{tikzpicture}[spy using outlines={rectangle, magnification=3, size= 42.5pt}]
\node[anchor=south west, inner sep=0] at (0,0){\includegraphics[height=90pt]{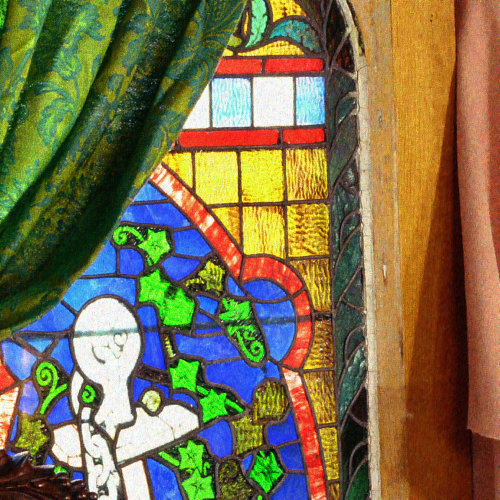}};
\spy [blue] on (0.8,2.3) in node [right] at (0.02,-30pt);
\spy [red] on (1.2,1.4) in node [right] at (1.68,-30pt);
\end{tikzpicture}
\captionsetup{justification=centering}
\caption*{ZS-N2N\\\textbf{27.6}/0.060}
\end{minipage}
\caption{Visual comparison of image \#1 in McMaster18 \cite{zhang2011color} for Gaussian noise level $\sigma = 10/255$ and Poissonian noise with peak intensity 100.
\textbf{Bold} and \underline{underline} indicate the best and second best results, respectively.}
\label{fig:zoominP1}
\end{center}
\end{figure*}

\begin{figure*}
\begin{center}
\begin{minipage}{0.19\linewidth}
\centering
\begin{tikzpicture}[spy using outlines={rectangle, magnification=3, size= 42.5pt}]
\node[anchor=south west, inner sep=0] at (0,0){\includegraphics[height=90pt]{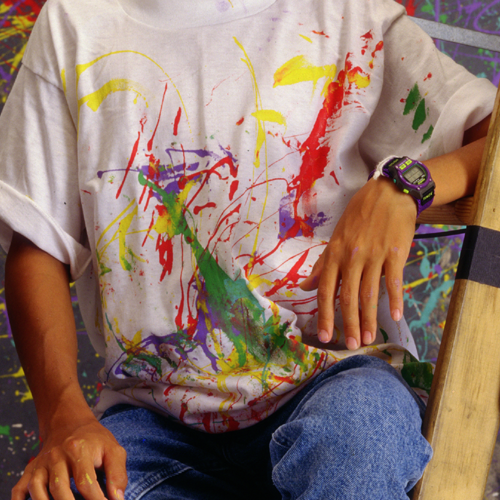}};
\spy [blue] on (0.8,2.3) in node [right] at (0.02,-30pt);
\spy [red] on (1.2,1.4) in node [right] at (1.68,-30pt);
\end{tikzpicture}
\captionsetup{justification=centering}
\caption*{Ground Truth\\PSNR/LPIPS}
\end{minipage}
\begin{minipage}{0.19\linewidth}
\centering
\begin{tikzpicture}[spy using outlines={rectangle, magnification=3, size= 42.5pt}]
\node[anchor=south west, inner sep=0] at (0,0){\includegraphics[height=90pt]{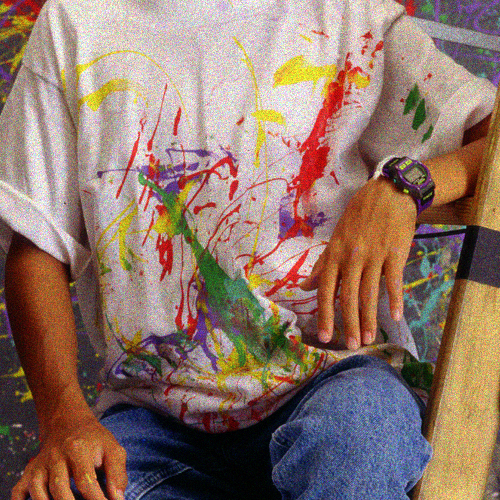}};
\spy [blue] on (0.8,2.3) in node [right] at (0.02,-30pt);
\spy [red] on (1.2,1.4) in node [right] at (1.68,-30pt);
\end{tikzpicture}
\captionsetup{justification=centering}
\caption*{Noisy\\23.3/0.278}
\end{minipage}
\begin{minipage}{0.19\linewidth}
\centering
\begin{tikzpicture}[spy using outlines={rectangle, magnification=3, size= 42.5pt}]
\node[anchor=south west, inner sep=0] at (0,0){\includegraphics[height=90pt]{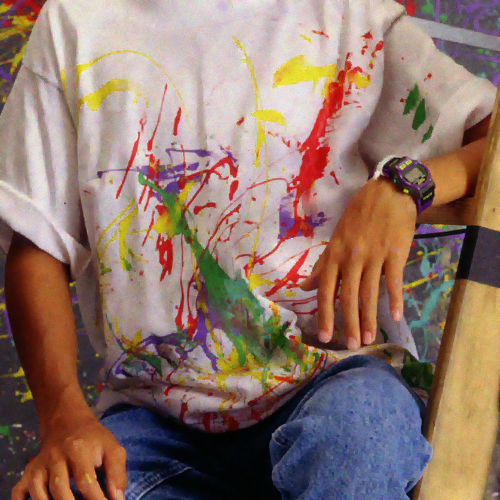}};
\spy [blue] on (0.8,2.3) in node [right] at (0.02,-30pt);
\spy [red] on (1.2,1.4) in node [right] at (1.68,-30pt);
\end{tikzpicture}
\captionsetup{justification=centering}
\caption*{\textbf{Proposed}\\30.7/\textbf{0.062}}
\end{minipage}
\begin{minipage}{0.19\linewidth}
\centering
\begin{tikzpicture}[spy using outlines={rectangle, magnification=3, size= 42.5pt}]
\node[anchor=south west, inner sep=0] at (0,0){\includegraphics[height=90pt]{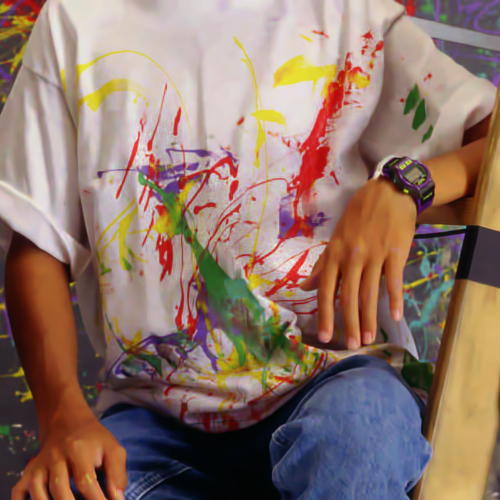}};
\spy [blue] on (0.8,2.3) in node [right] at (0.02,-30pt);
\spy [red] on (1.2,1.4) in node [right] at (1.68,-30pt);
\end{tikzpicture}
\captionsetup{justification=centering}
\caption*{BM3D\\\textbf{31.6}/0.141}
\end{minipage}
\begin{minipage}{0.19\linewidth}
\centering
\begin{tikzpicture}[spy using outlines={rectangle, magnification=3, size= 42.5pt}]
\node[anchor=south west, inner sep=0] at (0,0){\includegraphics[height=90pt]{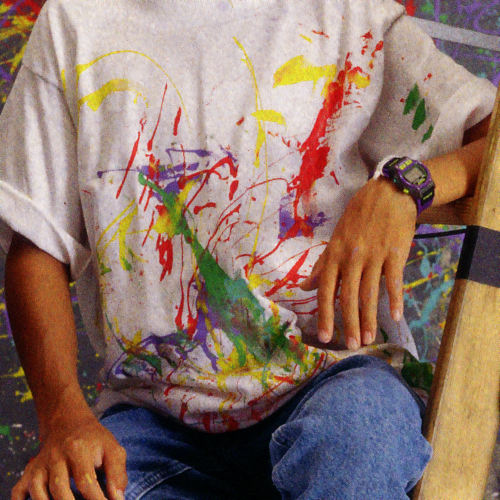}};
\spy [blue] on (0.8,2.3) in node [right] at (0.02,-30pt);
\spy [red] on (1.2,1.4) in node [right] at (1.68,-30pt);
\end{tikzpicture}
\captionsetup{justification=centering}
\caption*{ZS-N2N\\30.4/0.097}
\end{minipage}
\caption{Visual comparison of image \#5 in McMaster18 \cite{zhang2011color} for Gaussian noise level $\sigma = 20/255$ and Poissonian noise with peak intensity 100.
\textbf{Bold} and \underline{underline} indicate the best and second best results, respectively.}
\label{fig:zoominP2}
\end{center}
\end{figure*}

\begin{figure*}
\begin{center}
\begin{minipage}{0.19\linewidth}
\centering
\begin{tikzpicture}[spy using outlines={rectangle, magnification=3, size= 42.5pt}]
\node[anchor=south west, inner sep=0] at (0,0){\includegraphics[height=90pt]{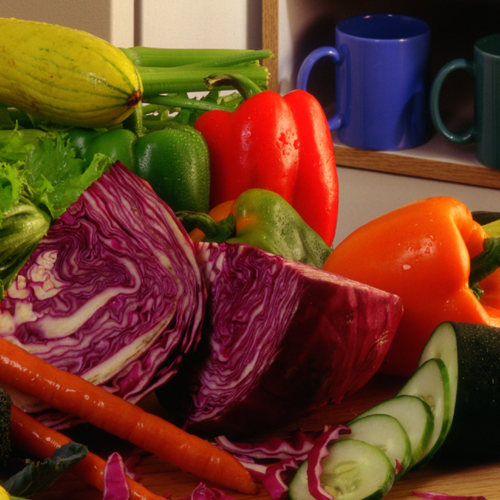}};
\spy [blue] on (0.8,2.3) in node [right] at (0.02,-30pt);
\spy [red] on (1.2,1.4) in node [right] at (1.68,-30pt);
\end{tikzpicture}
\captionsetup{justification=centering}
\caption*{Ground Truth\\PSNR/LPIPS}
\end{minipage}
\begin{minipage}{0.19\linewidth}
\centering
\begin{tikzpicture}[spy using outlines={rectangle, magnification=3, size= 42.5pt}]
\node[anchor=south west, inner sep=0] at (0,0){\includegraphics[height=90pt]{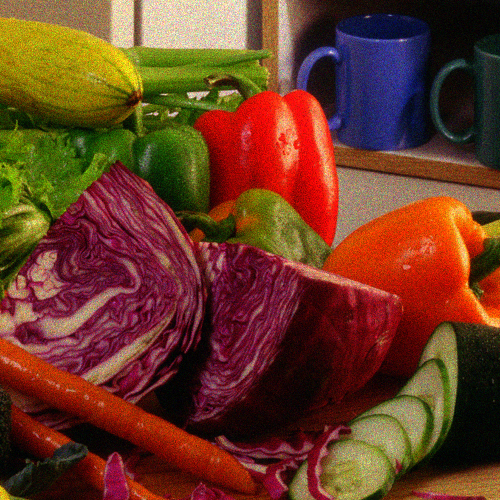}};
\spy [blue] on (0.8,2.3) in node [right] at (0.02,-30pt);
\spy [red] on (1.2,1.4) in node [right] at (1.68,-30pt);
\end{tikzpicture}
\captionsetup{justification=centering}
\caption*{Noisy\\25.9/0.250}
\end{minipage}
\begin{minipage}{0.19\linewidth}
\centering
\begin{tikzpicture}[spy using outlines={rectangle, magnification=3, size= 42.5pt}]
\node[anchor=south west, inner sep=0] at (0,0){\includegraphics[height=90pt]{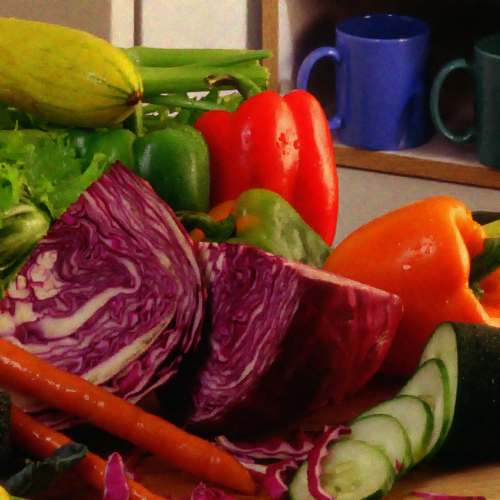}};
\spy [blue] on (0.8,2.3) in node [right] at (0.02,-30pt);
\spy [red] on (1.2,1.4) in node [right] at (1.68,-30pt);
\end{tikzpicture}
\captionsetup{justification=centering}
\caption*{\textbf{Proposed}\\33.1/\textbf{0.043}}
\end{minipage}
\begin{minipage}{0.19\linewidth}
\centering
\begin{tikzpicture}[spy using outlines={rectangle, magnification=3, size= 42.5pt}]
\node[anchor=south west, inner sep=0] at (0,0){\includegraphics[height=90pt]{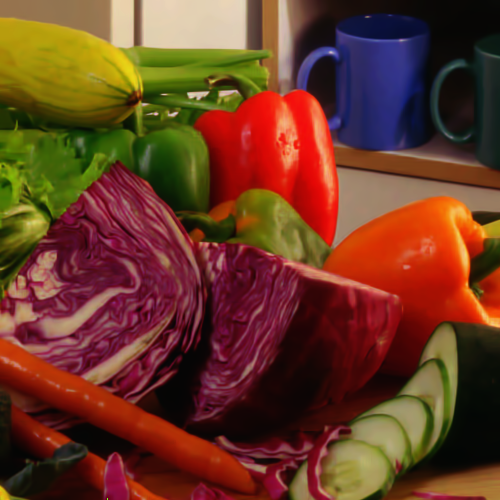}};
\spy [blue] on (0.8,2.3) in node [right] at (0.02,-30pt);
\spy [red] on (1.2,1.4) in node [right] at (1.68,-30pt);
\end{tikzpicture}
\captionsetup{justification=centering}
\caption*{BM3D\\\textbf{34.0}/0.099}
\end{minipage}
\begin{minipage}{0.19\linewidth}
\centering
\begin{tikzpicture}[spy using outlines={rectangle, magnification=3, size= 42.5pt}]
\node[anchor=south west, inner sep=0] at (0,0){\includegraphics[height=90pt]{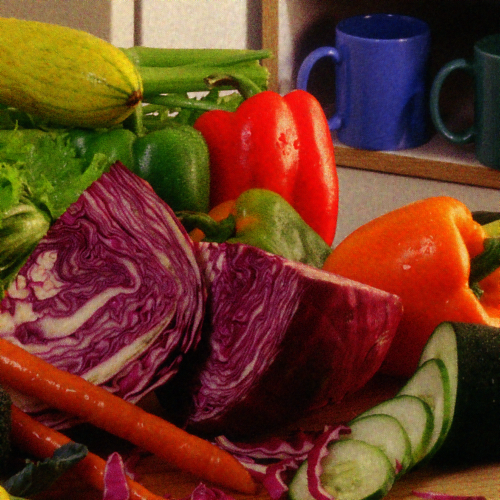}};
\spy [blue] on (0.8,2.3) in node [right] at (0.02,-30pt);
\spy [red] on (1.2,1.4) in node [right] at (1.68,-30pt);
\end{tikzpicture}
\captionsetup{justification=centering}
\caption*{ZS-N2N\\32.7/0.098}
\end{minipage}
\caption{Visual comparison of image \#10 in McMaster18 \cite{zhang2011color} for Gaussian noise level $\sigma = 20/255$ and Poissonian noise with peak intensity 100.
\textbf{Bold} and \underline{underline} indicate the best and second best results, respectively.}
\label{fig:zoominP2}
\end{center}
\end{figure*}

\subsection{Contrast Enhancement}
As discussed in Section~\ref{sec:Contrast}, the function $\varphi(u_n)$ should be symmetric, concave for $u_n\geq 0$, 
and have slope $>1$ at $u_n=0$.
There are several choices of function $\varphi$, like hyperbolic tangent function,
$\varphi(u_n) = \alpha \tanh(\beta u_n)$ with $\beta>1$. 
Here we choose another function that is modified with gamma correction as follows:

\begin{equation}
\varphi(u_n) = 
\begin{cases}
\sgn (u_n) \lambda^{\gamma-1} |u_n|, & |u_n| < \lambda, \\
\sgn (u_n) |u_n|^{\gamma},  &|u_n| \geq \lambda .
\end{cases}
\end{equation}

\begin{figure*}
\centering
\begin{minipage}{0.24\linewidth}
\centering
\includegraphics[height = 120pt]{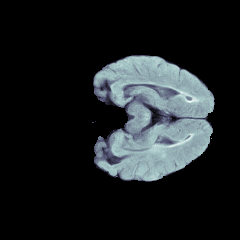}
\captionsetup{justification=centering}
\caption*{Given}
\end{minipage}
\begin{minipage}{0.24\linewidth}
\centering
\includegraphics[height = 120pt]{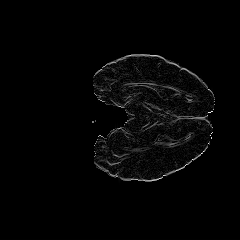}
\captionsetup{justification=centering}
\caption*{Column Edge}
\end{minipage}
\begin{minipage}{0.24\linewidth}
\centering
\includegraphics[height = 120pt]{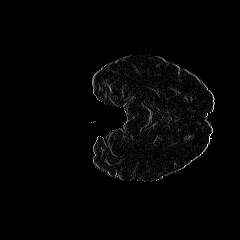}
\captionsetup{justification=centering}
\caption*{Row Edge}
\end{minipage}
\begin{minipage}{0.24\linewidth}
\centering
\includegraphics[height = 120pt]{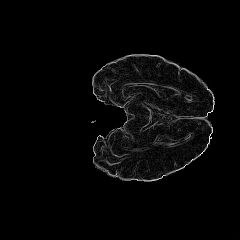}
\captionsetup{justification=centering}
\caption*{Edge}
\end{minipage}
\begin{minipage}{0.24\linewidth}
\vspace{2ex}
\centering
\includegraphics[height = 120pt]{figures/brain60/truth.png}
\captionsetup{justification=centering}
\caption*{Given}
\end{minipage}
\begin{minipage}{0.24\linewidth}
\vspace{2ex}
\centering
\includegraphics[height = 120pt]{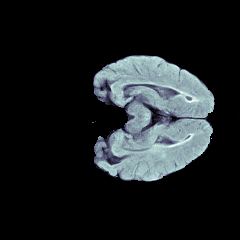}
\captionsetup{justification=centering}
\caption*{Enhanced $\gamma = 0.7$}
\end{minipage}
\begin{minipage}{0.24\linewidth}
\vspace{2ex}
\centering
\includegraphics[height = 120pt]{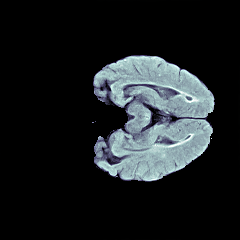}
\captionsetup{justification=centering}
\caption*{Enhanced $\gamma = 0.5$}
\end{minipage}
\begin{minipage}{0.24\linewidth}
\vspace{2ex}
\centering
\includegraphics[height =120pt]{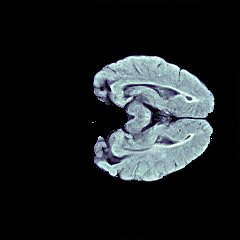}
\captionsetup{justification=centering}
\caption*{Enhanced $\gamma = 0.3$}
\end{minipage}
\caption{
Experiments of local contrast enhancement.
The image is a slice of brain MRI scan from BraTS 2018 \cite{menze2014multimodal, bakas2017advancing, bakas2018identifying}.
}
\label{fig:edge1}
\end{figure*}

The contrast enhancement and edge detection are shown in \Fig{fig:edge1}.
Here we choose $\lambda = 0.5$ and $\gamma = 0.7, 0.5, 0.3$ to elaborate
different contrast enhancement effects.

\subsection{Inpainting}
\begin{figure*}
\centering
\begin{minipage}{0.24\linewidth}
\centering
\includegraphics[height = 120pt]{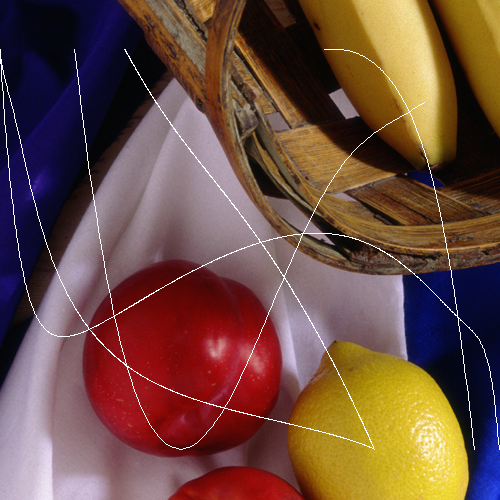}
\captionsetup{justification=centering}
\caption*{Given (1-2 pixels width)}
\end{minipage}
\begin{minipage}{0.24\linewidth}
\centering
\includegraphics[height = 120pt]{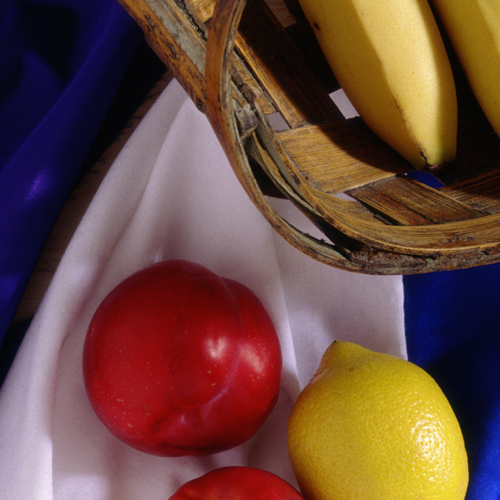}
\captionsetup{justification=centering}
\caption*{Inpainted}
\end{minipage}
\begin{minipage}{0.24\linewidth}
\centering
\includegraphics[height = 120pt]{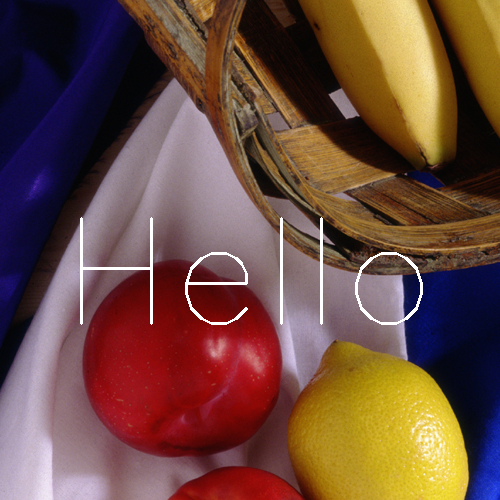}
\captionsetup{justification=centering}
\caption*{Given (2-3 pixels width)}
\end{minipage}
\begin{minipage}{0.24\linewidth}
\centering
\includegraphics[height = 120pt]{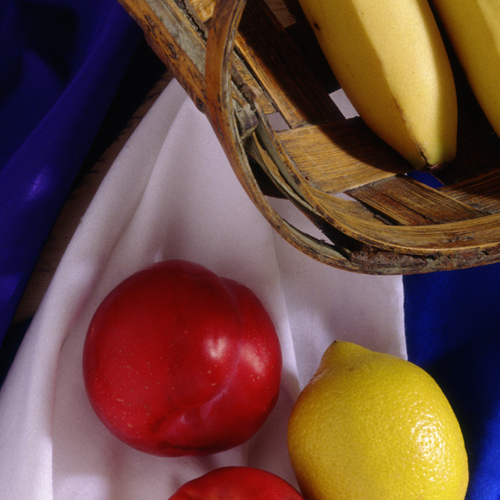}
\captionsetup{justification=centering}
\caption*{Inpainted}
\end{minipage}
\begin{minipage}{0.24\linewidth}
\vspace{2ex}
\centering
\includegraphics[height = 120pt]{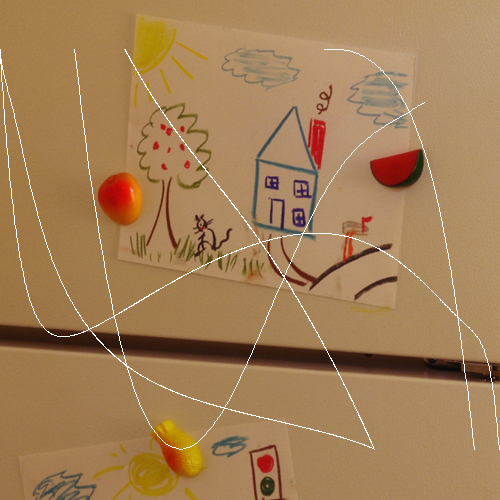}
\captionsetup{justification=centering}
\caption*{Given (1-2 pixels width)}
\end{minipage}
\begin{minipage}{0.24\linewidth}
\vspace{2ex}
\centering
\includegraphics[height = 120pt]{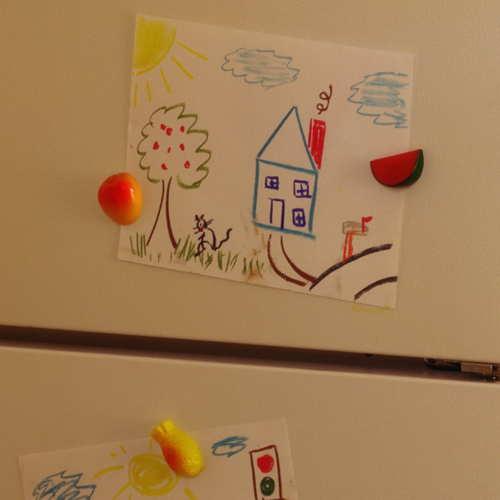}
\captionsetup{justification=centering}
\caption*{Inpainted}
\end{minipage}
\begin{minipage}{0.24\linewidth}
\vspace{2ex}
\centering
\includegraphics[height = 120pt]{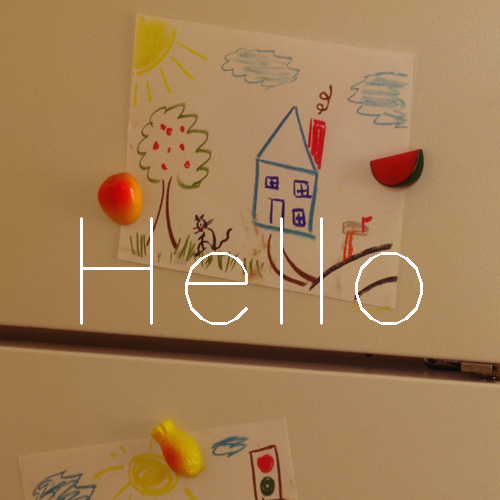}
\captionsetup{justification=centering}
\caption*{Given (2-3 pixels width)}
\end{minipage}
\begin{minipage}{0.24\linewidth}
\vspace{2ex}
\centering
\includegraphics[height = 120pt]{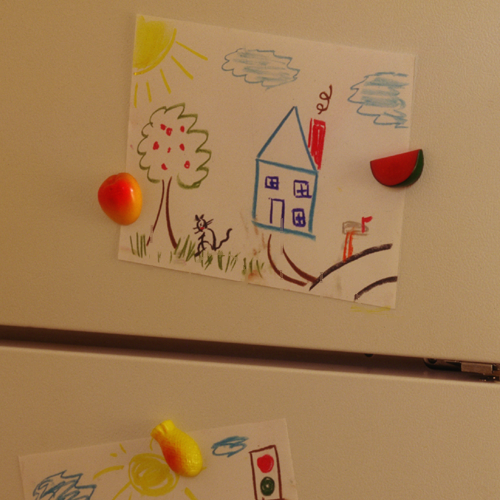}
\captionsetup{justification=centering}
\caption*{Inpainted}
\end{minipage}
\caption{
Experiments of image inpainting.
The images are \#13 and \#14 in McMaster18 \cite{zhang2011color}.
}
\label{fig:inpaint}
\end{figure*}

The model can easily adapted to inpainting problem by simply setting the parameter $\sigma_Z^2$
to be large enough (e.g. $\sigma_Z^2 = 100$) for the area (pixels) that need to be inpainted. \Fig{fig:inpaint} shows the inpainted
results of image with words (2-3 pixels width) and random white scratches (1-2 pixels width).

\subsection{Pseudocode of the Algorithm (steps 2-4) in Section~\ref{sec:BasicAlgorithm}}
\begin{algorithm}
\caption*{Forward-backward message passing \cite{loeliger2007factor, loeliger2016sparsity} for one row or one column}\label{alg:forback}
\textbf{Input:} $\displaystyle \msgb{W}{R_n''}, \msgb{\xi}{R_n''}$ for $n = 2, 3, \cdots, N$\\
\textbf{Output:} $\displaystyle r_n$ for $n = 2, 3, \cdots, N$, $\displaystyle \sigma_{\Delta_n}^2$ for $n = 3, 4, \cdots, N$
\begin{algorithmic}[1]\onehalfspacing
\State{\emph{Forward recursion:}}
\State{\emph{Initialize:} $\displaystyle \msgf{V}{R_2} \gets \msgb{W}{R_2'}^{-1}, \msgf{m}{R_2} \gets \msgb{W}{R_2''}^{-1} \msgb{\xi}{R_2''}$}
\For{$n = 2, 3, \cdots, N-1$}
    \State{$\displaystyle \msgf{V}{R_{n+1}'} \gets \msgf{V}{R_n} + \sigma_{\Delta_{n+1}}^2$}
    \State{$\displaystyle \msgf{m}{R_{n+1}'} \gets \msgf{m}{R_n}$}
    \State{$\displaystyle \msgf{W}{R_{n+1}} \gets \msgf{V}{R_{n+1}'}^{-1} + \msgb{W}{R_{n+1}''}$}
    \State{$\displaystyle \msgf{\xi}{R_{n+1}} \gets \msgf{V}{R_{n+1}'}^{-1} \msgf{m}{R_{n+1}'} + \msgb{\xi}{R_{n+1}''} $}
    \State{$\displaystyle \msgf{V}{R_{n+1}} \gets  \msgf{W}{R_{n+1}}^{-1}$}
    \State{$\displaystyle \msgf{m}{R_{n+1}} \gets  \msgf{W}{R_{n+1}}^{-1} \msgf{\xi}{R_{n+1}}$}
\EndFor
\State{\emph{Backward recursion:}}
\State{\emph{Initialize:} $\displaystyle \msgb{W}{R_N} \gets 0, \msgb{\xi}{R_N} \gets 0$}
\For{$n = N, N-1, \cdots, 3$}
    \State{$\displaystyle \msgb{W}{R_{n}'} \gets \msgb{W}{R_n} + \msgb{W}{R_{n}''}$}
    \State{$\displaystyle \msgb{\xi}{R_{n}'} \gets \msgb{\xi}{R_n} + \msgb{\xi}{R_{n}''}$}
    \State{$\displaystyle \msgb{V}{R_{n-1}} \gets \msgb{W}{R_n'}^{-1} + \sigma_{\Delta_n}^2$}
    \State{$\displaystyle \msgb{m}{R_{n-1}} \gets \msgb{W}{R_{n}'}^{-1} \msgb{\xi}{R_{n}'} $}
    \State{$\displaystyle \msgb{W}{R_{n-1}} = \msgb{V}{R_{n-1}}^{-1}$}
    \State{$\displaystyle \msgb{\xi}{R_{n-1}} = \msgb{W}{R_{n-1}} \msgb{m}{R_{n-1}}$}
\EndFor
\State{\emph{Compute posterior mean:}}
\For{$n = 2, 3, \cdots, N$}
    \State{$\displaystyle r_n \gets \frac{\msgf{V}{R_{n}}^{-1}\msgf{m}{R_{n}} + \msgb{V}{R_{n}}^{-1}\msgb{m}{R_{n}}}{\msgf{V}{R_{n}}^{-1} + \msgb{V}{R_{n}}^{-1}}$}
\EndFor
\State{\emph{Update NUP parameters ($\theta_n$ is updated implicitly \cite{loeliger2023nup}):}}
\For{$n = 3, 4, \cdots, N$}
    \State{$\displaystyle \Delta_n \gets r_n - r_{n-1}$}
    \State{$\displaystyle \sigma_{\Delta_n}^2 \gets \frac{\| \Delta_n \|^{2-p}}{\beta_{\Delta} p}$}
\EndFor

\end{algorithmic}
\end{algorithm}

\clearpage
\clearpage

\bibliographystyle{IEEEtran} 
\bibliography{reference} 

\end{document}